\documentclass[aps, prb, reprint, amssymb, amsfonts, groupedaddress]{revtex4-1}

\pdfoutput=1

\usepackage{amsmath}
\usepackage{bm}
\usepackage{graphicx}
\usepackage[unicode]{hyperref}
\usepackage[latin1]{inputenc}
\usepackage{array}
\usepackage{upgreek}
\usepackage{color}

\begin{document}

\title{Feasibility of imaging using Boltzmann polarization in nuclear Magnetic Resonance Force Microscopy}

\author{M. de Wit}
\author{G. Welker}
\author{J.J.T. Wagenaar}
\author{F.G. Hoekstra}
\author{T.H. Oosterkamp}
\email{oosterkamp@physics.leidenuniv.nl}
	
\affiliation{Leiden Institute of Physics, Leiden University, PO Box 9504, 2300 RA Leiden, The Netherlands}

\date{\today}

\begin{abstract}
We report on Magnetic Resonance Force Microscopy measurements of the Boltzmann polarization of the nuclear spins in copper by detecting the frequency shift of a soft cantilever. We use the time-dependent solution of the Bloch equations to derive a concise equation describing the effect of rf magnetic fields on both on- and off-resonant spins in high magnetic field gradients. We then apply this theory to saturation experiments performed on a 100 nm thick layer of copper, where we use the higher modes of the cantilever as source of the rf field. We demonstrate a detection volume sensitivity of only (40 nm)$^3$, corresponding to about 1.6$\cdot 10^4$ polarized copper nuclear spins. We propose an experiment on protons where, with the appropriate technical improvements, frequency-shift based magnetic resonance imaging with a resolution better than (10 nm)$^3$ could be possible. Achieving this resolution would make imaging based on the Boltzmann polarization competitive with the more traditional stochastic spin-fluctuation based imaging, with the possibility to work at milliKelvin temperatures.

\end{abstract}

\maketitle

\section{Introduction}
Magnetic Resonance Force Microscopy (MRFM) is a technique that combines magnetic resonance protocols with an ultrasensitive cantilever to measure the forces exerted by extremely small numbers of spins, with the immense potential of imaging biological samples with nanometer resolution \cite{zuger1993,rugar1994,sidles1995}. In the last 20 years, great steps have been taken towards this goal, with some milestones including the detection of a single electron spin \cite{rugar2004}, the magnetic resonance imaging of a tobacco mosaic virus with a spatial resolution of 4 nm \cite{degen2009}, and more recently the demonstration of a one-dimensional slice thickness below 2 nm for the imaging of a polystyrene film \cite{rose2018}. The experiments are typically performed by modulating the sample magnetization in resonance with the cantilever, and then measuring either the resulting change in the oscillation amplitude (force-based) or the frequency shift (force-gradient based). 

Both the force-based and force-gradient based experiments have some severe technical drawbacks, mainly associated to the cyclic inversion of the spin ensemble. For the coherent manipulation of the magnetization, alternating magnetic fields on the order of several mT are required \cite{poggio2007,nichol2012}. The dissipation associated with the generation of these fields is significant, and prevents experiments from being performed at milliKelvin temperatures, even for low duty-cycle MRMF protocols like cyclic-CERMIT \cite{garner2004,mamin2007}. Furthermore, the requirement that the magnetization is inverted continuously during the detection of the signal means only samples with a long rotating-frame spin-lattice relaxation time $T_{1\rho}$ are suitable.

For imaging of nuclei, previous experiments have almost exclusively focused on measuring the statistical polarization of the spin ensemble. However, the possibility to use the Boltzmann polarization instead would dramatically improve the efficiency of the measurement, as averaging $N$ times enhances the power signal-to-noise ratio (SNR) by a factor of $N$ for Boltzmann based measurements, compared to $\sqrt{N}$ for statistical polarization signals. There have been MRFM experiments based on the Boltzmann polarization, for instance in order to measure the relaxation times of nuclei \cite{garner2004,alexson2012,wagenaar2016}, but these experiment lacked the volume sensitivity required for imaging with a spatial resolution comparable to the statistical experiments.

In this work, we present measurements of the Boltzmann polarization of a copper sample at a temperature of 21 mK by detecting the frequency shift induced by a saturation experiment. We derive the time-dependent solution to the Bloch equations appropriate for typical MRFM experiments, obtaining a concise equation for the non-equilibrium response of both on- and off-resonant spins to a radio-frequent (rf) pulse. We apply this equation to show that we are able to measure the frequency shift of our resonator with a noise floor of 0.1 mHz. Furthermore, we demonstrate that we can use higher modes of the cantilever as the source of the alternating field in order to generate the required rf fields to saturate the magnetization of the spins with minimal dissipation \cite{wagenaar2017}. These results suggest that imaging based on the Boltzmann polarization could be possible, allowing for the first MRFM imaging experiments performed at milliKelvin temperatures down to 10 mK and using the magnet-on-tip geometry, as opposed to the sample-on-tip geometry more commonly found. We substantiate this claim by using the specifications of the current experiments to calculate the resolution for an imaging experiment on protons based on measuring the Boltzmann polarization.

\section{Methods}

\subsection{Experimental setup}
\begin{figure}
	\centering
	\includegraphics[width=\columnwidth]{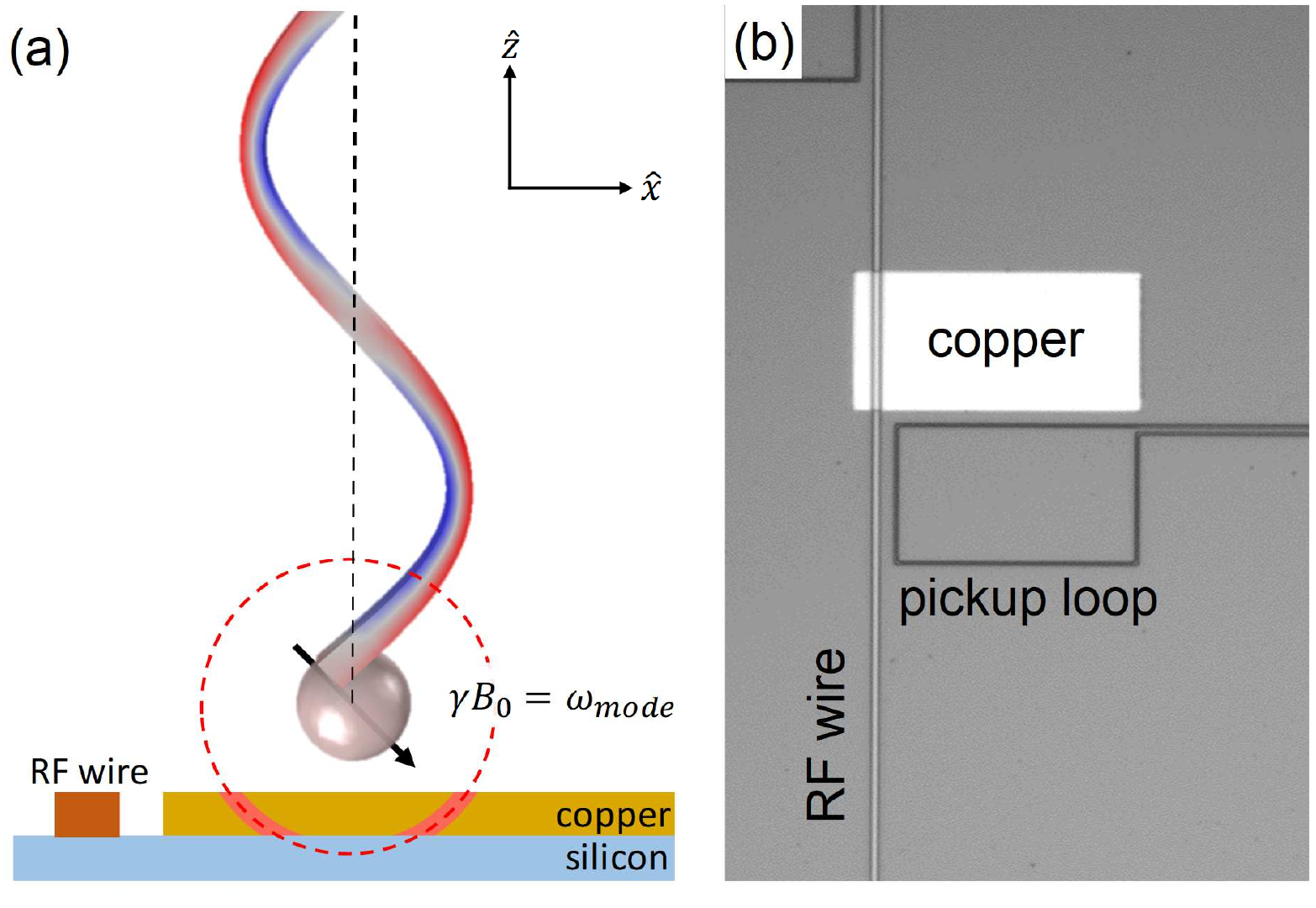}
	\caption{(a) schematic of the measurement setup. A rf wire is used to generate a rf field $B_{rf}$ directly, or to excite higher modes of the cantilever to generate $B_{rf}$ fields with minimal dissipation. The rf pulse removes the Boltzmann polarization of spins located within an near the resonance slice (red region), inducing mHz shifts of the cantilevers fundamental resonance frequency at 3.0 kHz. (b) optical microscope image of the detection chip, showing the NbTiN pickup loop and rf wire, and the copper sample with a thickness of 100 nm.}
	\label{figure:Setup}
\end{figure}

We improve on earlier measurements in our group on the nuclear spins in a copper sample. The setup and measurement procedure strongly resemble those used in that previous work \cite{wagenaar2016}. The operating principle of the setup is shown in Fig. \ref{figure:Setup}(a). The heart of the setup is a soft single-crystal silicon cantilever (spring constant $k_0$ = 70 $\upmu$Nm$^{-1}$)\cite{chui2003} with at the end a magnetic particle with a radius $R_0$ = 1.7 $\upmu$m, resulting in a natural resonance frequency $f_0 = \omega_0 / (2\pi) \sim$ 3.0 kHz, an intrinsic Q-factor $Q_0 \sim 3 \cdot 10^4$, and a thermal force noise at 20 mK of 0.4 aN/$\sqrt{\mathrm{Hz}}$. The magnet induces a static magnetic field $B_0$ which can be well approximated by the field of a perfect magnetic dipole. The field strength reduces quickly as the distance to the center of the magnet increases, but for typical experimental parameters is of the order of a few hundred mT. When the cantilever is placed at a height $h$ above a sample, spins in the sample couple to the resonator via the magnetic field gradient, inducing a frequency shift (see Sec. \ref{sec:freq_shift}). A rf pulse with frequency $\omega_{rf}$ can be used to remove the polarization of the spins that are resonant with this pulse, i.e. the spins that are within the resonance slice where $|B_0| = \omega_{rf}/\gamma$, with $\gamma$ the gyromagnetic ratio of the spins (in Fig. \ref{figure:Setup}(a) the resonant slice is marked in red). We will refer to this procedure as a saturation experiment or saturation pulse. The theoretical background of the saturation experiment is given in Sec. \ref{sec:bloch}.

Our particular MRFM setup is designed to be operated at temperatures close to 10 mK in order to decrease the thermal noise in the cantilever and increase the Boltzmann polarization of the sample. In order to do this, we have developed a detection scheme without a laser which is based on a SQUID detection \cite{usenko2011}. In this scheme we measure the flux from the moving magnetic particle using a pickup loop (see Fig. \ref{figure:Setup}(b)) connected via a gradiometric transformer to the input coil of a DC SQUID \footnote{Magnicon GMBH. Integrated 2-stage current sensor, type C70M116W}. Additionally, we use a superconducting NbTiN rf wire to send rf currents to the sample \cite{thoen2017}. The MRFM setup is mounted at the bottom of a mechanical vibration isolation stage, and the cryostat has been modified to reduce vibrations originating from the pulse tube refrigerator \cite{haan2014}.

The rf pulse can be applied using two methods, both shown in Fig. \ref{figure:Setup}(a). First of all, we can use a rf wire to send an alternating current which generates a magnetic field directly. This allows for precise control of the pulse shape and amplitude, but at the cost of some heating of the sample due to AC dissipation in the superconducting rf wire. The amplitude of the rf field $B_{rf}$ is inversely proportional to the distance to the rf wire, dictating that all measurements have to be done as close to the rf wire as possible (preferably within several micrometers). At a distance of 5 $\upmu$m from the rf wire, we can generate magnetic fields (in the rotating frame of the spins, see Sec. \ref{sec:bloch}) of up to 0.3 mT. An alternative method to generate the required rf field is by using the higher modes of the cantilever, the proof of concept of which was recently demonstrated by \citeauthor{wagenaar2017}\cite{wagenaar2017}. Generating rf fields using the higher modes can be done with a small current in the rf wire to generate a magnetic drive field, or by using a piezo at the base of the cantilever, allowing experiments at larger distances from the rf wire, or even without one. In our experiment, we use a small current in the rf wire (on the order of $\sim$ 10 $\upmu$A) to excite one of the higher modes of the cantilever, as illustrated in Fig. \ref{figure:Setup}(a). The motion of the higher mode induces a small rotation of the magnet, which results in the generation of an amplified $B_{rf}$ at the frequency of the excited higher mode perpendicular to the tip field. In this way, rf fields can be generated with negligible dissipation.
 
The copper sample used in the experiment is patterned on the detection chip close to both the rf wire and the pickup loop, as shown in Fig. \ref{figure:Setup}(b). The copper sample is a sputtered film with a thickness of 100 nm, capped with a 20 nm layer of gold to prevent oxidation. The thickness of the sample was chosen to be 100 nm in order to reduce eddy current in the copper which deteriorate the Q-factor of the cantilever and thereby the measurement sensitivity (for metal films with a thickness less than the skin depth, eddy current dissipation scales with the cube of the thickness \cite{meyer1989}). The copper overlaps with the rf wire in order to give the sample a well defined potential. Besides the thermal conductance of the silicon substrate, there is no additional thermalization used to cool the copper. The cantilever can be positioned above the copper with a lateral accuracy of several micrometers, and should be as close as possible to both the rf wire and the edge of the pickup loop.

Copper was selected as a sample for its favorable NMR properties for a MRFM experiment, especially the long $T_1$ relaxation times of the order of 1 to 100 s at low temperatures resulting from the Korringa relation\cite{korringa1950}. All relevant NMR properties of both copper isotopes can be found in Table. \ref{table:copper}.

\begin{table}
		\begin{tabular}{c c c c c}
		\hline\hline
		Parameter	& Variable & $^{63}$Cu	& $^{65}$Cu	\\	 
		\hline
		Spin 	& $S$		& 3/2   	& 3/2 \\
		Natural abundance 	& 		& 69 \%   	& 31 \% \\
		Gyromagnetic Ratio 	& $\gamma /(2\pi)$		& 11.3 MHz/T   	& 12.1 MHz/T \\
		spin-spin relaxation time	&	$T_2$	&	0.15 ms	&	0.15 ms	\\
		\hline\hline 
		\end{tabular}
	\caption{Overview of the relevant NMR parameters for the two isotopes of copper. We assume a combined spin density $\rho$ = 85 spins/nm$^3$, and spin-lattice relaxation times $T_1$ dictated by the Korringa relation $T T_1$ = 1.15 sK \citep{lounasmaa1997,oja1997,pobell2007}.} 
	\label{table:copper}
\end{table}

\begin{figure}
\centering
\includegraphics[width=\columnwidth]{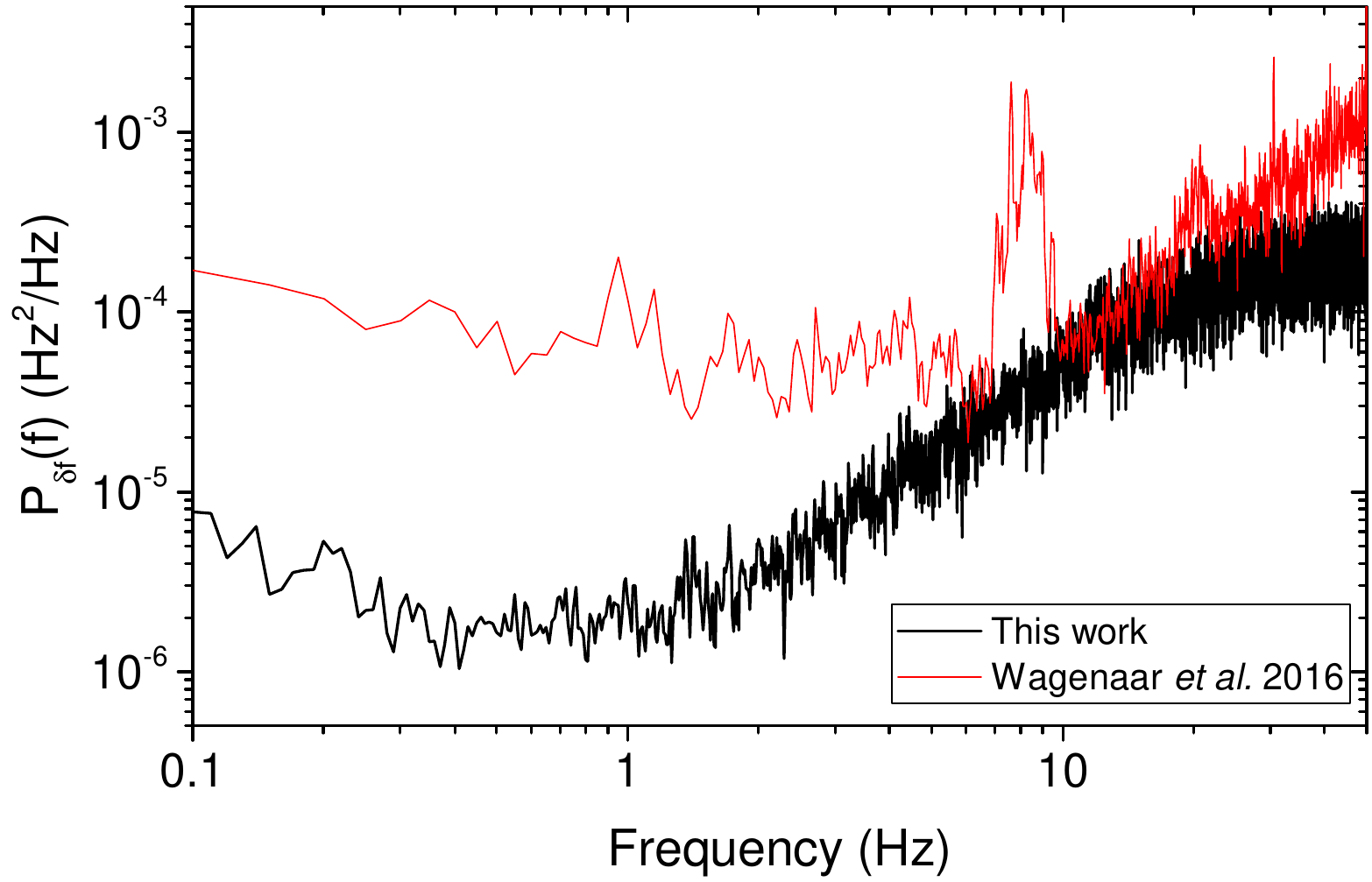}
\caption{Frequency noise spectrum $P_{\delta f}$ measured at a height of 1.3 $\upmu$m. In red we see the frequency noise spectrum from the initial experiment in our group, measured with a cantilever amplitude of 110 nm\textsubscript{rms}\cite{wagenaar2016}. In black we see the current experiment, measured with a cantilever amplitude of 43 nm\textsubscript{rms}. The noise floor has been reduced by almost 2 orders of magnitude.}
\label{figure:PLL_Noise}
\end{figure}

We have employed a series of improvements to the setup to enhance the frequency noise floor of the measurement, and thus increase the sensitivity. The improvement is obvious when looking at the noise spectrum of the frequency, as shown in Fig. \ref{figure:PLL_Noise}. The spectrum is measured by driving the cantilever with an amplitude of 43 nm\textsubscript{rms} and tracking the resonance frequency using a phase-locked loop (PLL) of a Zurich Instruments lock-in amplifier with a detection bandwidth of 40 Hz. The PLL feedback signal is sent to a spectrum analyzer. In black we see the frequency noise spectrum of the current setup, while in red we see that from the experiment from 2016 on a 300 nm thick copper film performed in our group \cite{wagenaar2016}. Both spectra were measured at a height of 1.3 $\upmu$m above a copper sample. The total frequency noise is given by the sum of the thermal noise, the detection noise, and 1/f noise typically attributed to the sample \cite{isaac2016,wagenaar2016}:
\begin{equation}
P_{\delta f} (f) = P_{\delta f}^{thermal} + P^{det} f^2 + P^{sample} f^{-1}
\end{equation}
The noise reduction of nearly 2 orders of magnitude is due to a combination of several technical improvements. Improved vibration isolation and cantilever thermalization have reduced the thermodynamic temperature of the cantilever from 132 mK to less than 50 mK. An improved design of the pickup loop resulted in an amplitude detection noise floor of 30 pm/$\sqrt{\mathrm{Hz}}$, determined from the measured transfer between the cantilever motion and the SQUID's output voltage. This allows for a much lower cantilever drive amplitude with the same detection frequency noise. The biggest improvement seems to be the reduction of the thickness of the copper film. Because the dissipated power of the eddy currents in the film scales strongly with the thickness of the film, we find that the Q-factor has increased from 317 for the 300 nm film to almost 5000 for the 100 nm film. This reduces all three contribution to the frequency noise, particularly the 1/f noise which is mainly attributed to eddy currents in the sample. The thermal noise floor using these parameters is estimated to be 0.7 mHz/$\sqrt{\mathrm{Hz}}$, so the data in Fig. \ref{figure:PLL_Noise} is not thermally limited. With a 1 Hz detection bandwidth, the integrated frequency noise is as low as 1.8 mHz.

\subsection{Measurement procedure}
\begin{figure}
\centering
\includegraphics[width=\columnwidth]{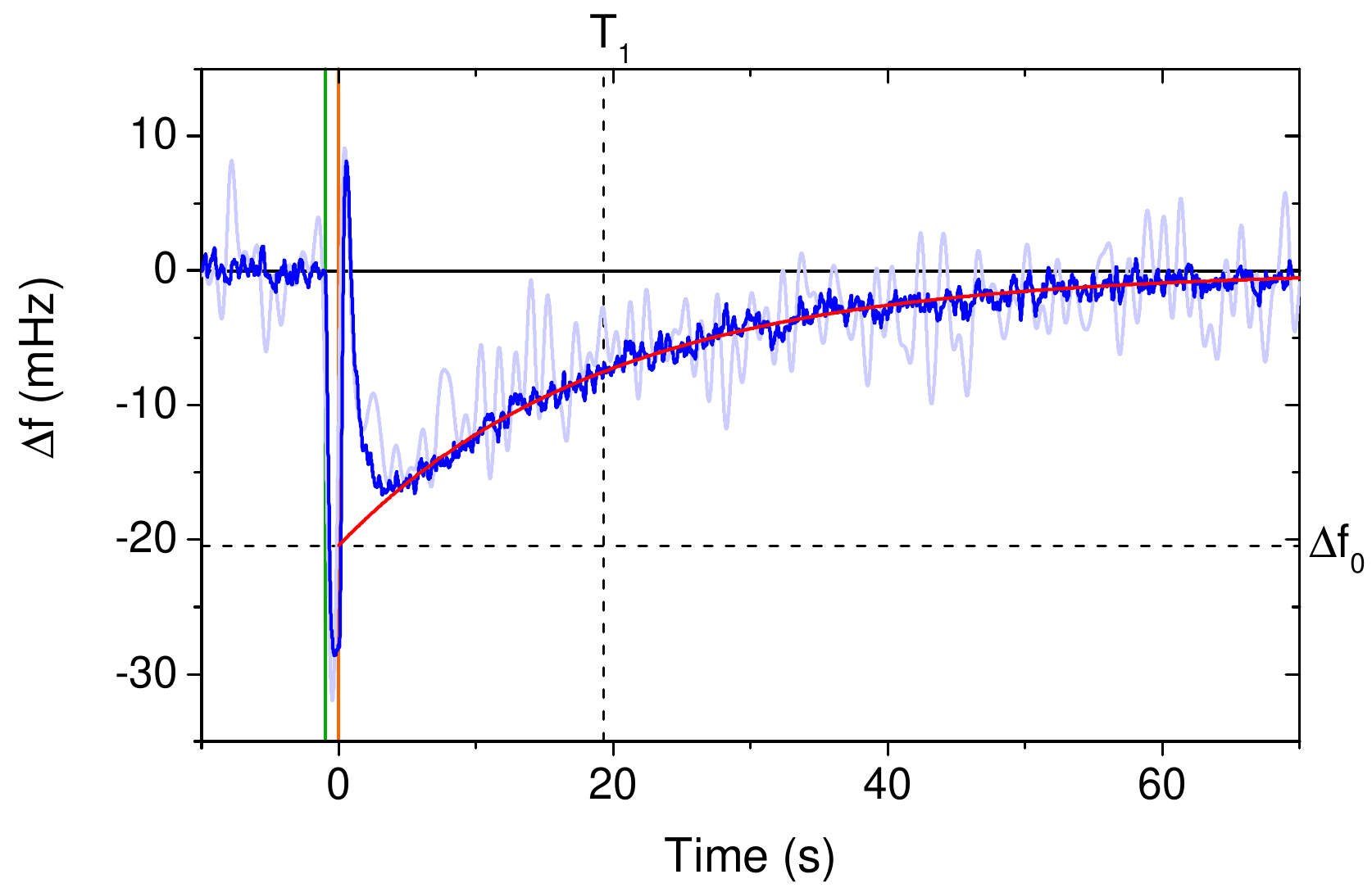}
\caption{Example of a typical measurement (at $T$ = 40 mK) where we show the frequency shift $\Delta f$ with respect to the equilibrium frequency $f_0$. The light blue line shows a single measurement of the frequency shift (after a 1 Hz low-pass filter). The dark blue line shows 50 averages. The red solid line is an exponential fit to the data following Eq. \ref{eq:exponential}, from which we extract the spin-lattice relaxation time $T_1$ and the frequency shift directly after the end of the pulse $\Delta f_0$. The green and orange vertical lines indicate the start and end of the saturation pulse.}
\label{figure:PLL_Example}
\end{figure}

A typical saturation recovery measurement (performed at a temperature $T$ = 40 mK) is shown in Fig. \ref{figure:PLL_Example}. Again a PLL is used to measure the frequency shift $\Delta f = f(t) - f_0$. A 1 Hz low-pass filter is used to remove the effects of the detection noise visible in Fig. \ref{figure:PLL_Noise}. Before the first rf pulse, $f_0$ is determined by measuring the cantilever resonance frequency in equilibrium. Then a rf pulse with a certain duration $t_p$ and strength $B_{rf}$ is turned on. In the figure, the start and end are indicated by the green and orange vertical lines. It is optional to switch off the PLL and cantilever drive during the pulse to reduce the oscillation amplitude and the associated broadening of the resonant slice. During the pulse, we observe frequency shifts which we attribute to a combination of electrostatic effects and slight local heating of the sample. After the pulse, the frequency shift relative to $f_0$ is measured. The obtained recovery curve can be fitted to
\begin{equation} \label{eq:exponential}
	\Delta f(t) = \Delta f_0 ~ e^{- \left( t-t_0 \right) / T_1},
\end{equation}
with $\Delta f_0$ the direct frequency shift at the time of the end of the pulse $t_0$. The light blue curve in Fig. \ref{figure:PLL_Example} shows the result of a single measurement of the frequency shift (with a 1 Hz low-pass filter), the dark blue curve shows the result of 50 averages. In red we show the best fit to the data using equation \ref{eq:exponential}.

\subsection{Spin dynamics in MRFM} \label{sec:bloch}
In order to fully understand the observed frequency shifts, we need to find the final magnetization of the spins coupled to the magnetic field of our cantilever after a saturation pulse. The behaviour of spins in alternating magnetic fields is well understood from conventional NMR, but the analysis is often limited to steady-state solutions \cite{abragam1961}. This limit works well for most NMR applications where the alternating fields are of sufficient strength and duration that the magnetization of the spin ensemble has reached an equilibrium duing the pulse, but this does not necessary work for MRFM due to the large magnetic field gradient and often weak oscillating magnetic fields. Therefore, we will derive equations for the time dependence of the magnetization of spins during a rf pulse, also for spins not meeting the resonance condition. These equations are then used to derive the effective resonance slice thickness in an MRFM experiment, a crucial component in trying to decrease the detection volume and thereby optimize the imaging resolution.

The time evolution of spins subjected to a large static magnetic field ($B_0$) and a small alternating magnetic field ($B_{rf}$) perpendicular to the static field has long been understood using the Bloch equations \cite{bloch1946}. In the rotating frame, the equations of motion of the magnetization $\bm{m}(t)$ subjected to an effective magnetic field $\bm{B}_{eff} = \left(B_0 - \omega / \gamma \right)\bm{\hat{k}}+B_{rf}\bm{\hat{i}}$ are given by
\begin{equation}
	\begin{aligned}
	\frac{d m_x}{dt}&=-\Delta\omega m_y-\frac{m_x(t)}{T_2}\\
	\frac{d m_y}{dt}&=\omega_{1}m_z+\Delta\omega m_x-\frac{m_y(t)}{T_2}\\
	\frac{d m_z}{dt}&=-\omega_{1} m_y-\frac{m_z(t)-m_0}{T_1}
	\end{aligned}
\label{eq:system}
\end{equation}
Here $\gamma$ is the gyromagnetic ratio of the spins, $T_1$ and $T_2$ the spin-lattice (longitudinal) and spin-spin (transverse) relaxation times, the detuning $\Delta\omega\equiv\omega-\omega_0$ with $\omega_0 = \gamma B_0$ the Larmor frequency, and $\omega_1\equiv \gamma B_{rf}$. $m_0$ is the initial magnetization in thermal equilibrium. To solve this system of differential equations, it is convenient to rewrite them in vector notation as
\begin{equation} \label{eq:vector}
	\dot{\bm{m}}=A\bm{m}+\bm{b},
\end{equation}
with the source term $\bm{b} = \frac{m_0}{T_1} \hat{k}$, and $A$ given by
\begin{align}
	A=
	\begin{pmatrix}
	 -\frac{1}{T_2}	& -\Delta\omega 	& 0 \\
	 \Delta\omega  	&  -\frac{1}{T_2} 	& \omega_1 \\
	  0 			& -\omega_1		 	& -\frac{1}{T_1}
	 \end{pmatrix}
\end{align}
The steady state solution is now easy to derive by solving the differential equation after setting $\dot{\bm{m}} = 0$. Note that $m_x$ and $m_y$ are rotating with the Larmor frequency around the z-axis. As the resonance frequency of the cantilevers used in MRFM are typically much lower than the Larmor frequency, any coupling of these two components to the cantilever averages out over time. Therefore, we are only interested in the z-component of the magnetization, which is the same in the rotating frame as in the laboratory frame \cite{abragam1961,slichter1990}:
\begin{equation}
\begin{aligned}
m_{z,\infty}&=\frac{1+\Delta\omega^2T_2^2}{1+\Delta\omega^2T_2^2+\omega_1^2T_1T_2}m_0\\
&\equiv p_{z} m_0
\end{aligned}
\label{eq:steadystate}
\end{equation}
In the last line we defined $p_{z}$ as the fraction of the magnetization that is removed by the $B_{rf}$ field if it is left on continuously.

In MRFM experiments the steady state solution described by Eq. \ref{eq:steadystate} is often not enough, as the rf pulses are not necessarily of sufficient strength and duration to fully saturate the magnetization of a spin ensemble. The time-dependent solution where $\dot{\bm{m}} \neq 0$ is given by the sum of the homogeneous solution ($\bm{b} = 0$) and the non-homogeneous steady state solution:
\begin{equation} \label{eq:4p18}
\begin{aligned}
m_z &= m_{z,\infty} + (m_0 - p_z m_0)e^{\lambda_z t} \\
&= p_z m_0 + (m_0 - p_z m_0) e^{-\frac{t}{T_1 p_z}},
\end{aligned}
\end{equation}
where $\lambda_z = 1/(T_1 p_z)$ is the third eigenvalue of the matrix $A$. Inserting this equation into Eq. \ref{eq:vector} confirms that it is a valid solution. The equation above gives the time-dependent z-magnetization of a spin ensemble after a rf magnetic field is turned on an left on. In deriving it, we have assumed that $T_2 \ll T_1$ and that the strength of the rf field is weak such that $\omega_1 T_2 \ll 1$. These assumptions give us a concise equation much more convenient for saturation experiments in MRFM than the expressions found in the general case \cite{mulkern1993,murase2011}.

\begin{figure}
\centering
\includegraphics[width=\columnwidth]{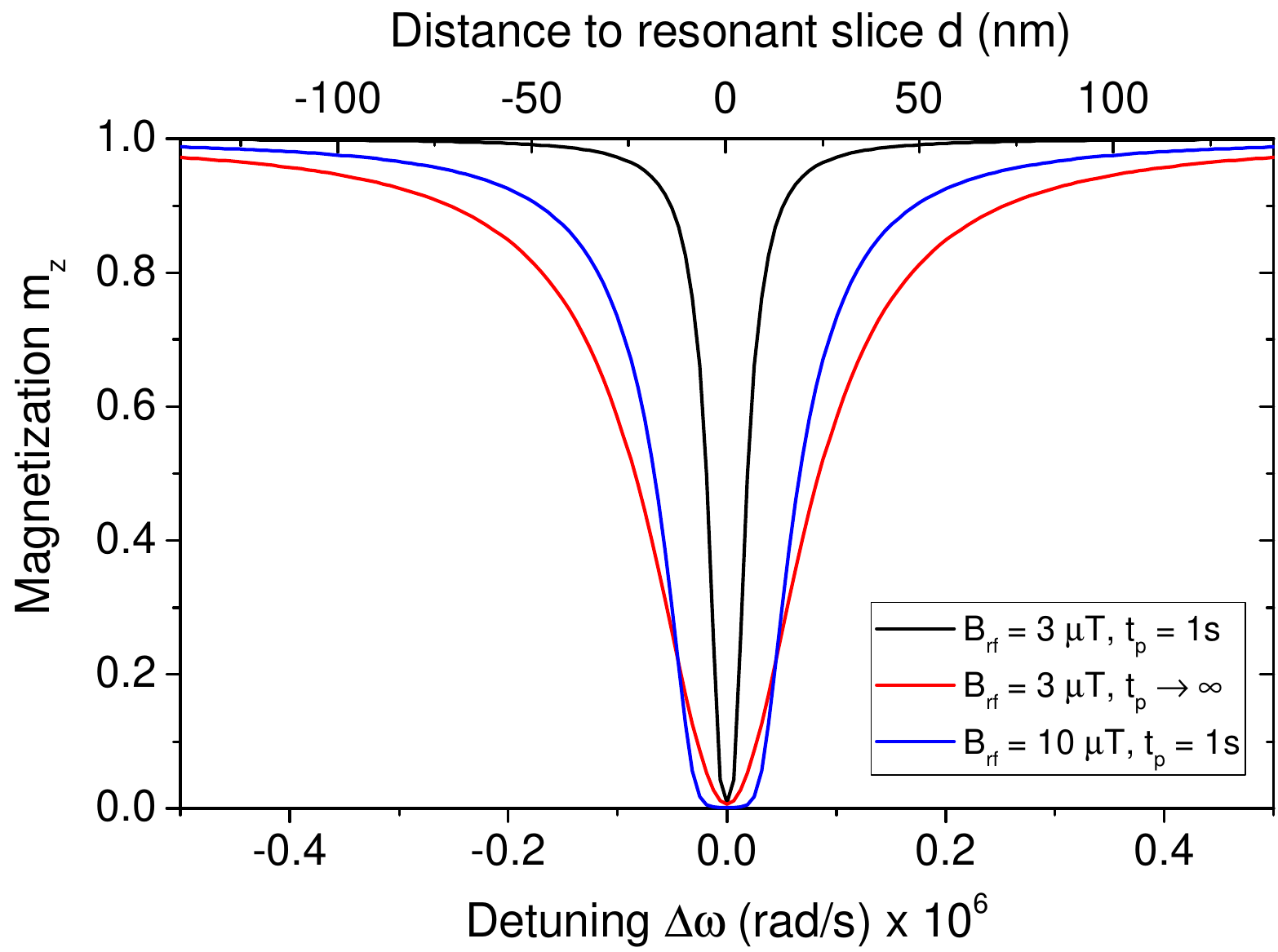}
\caption{Calculated magnetization $m_z$ after three different rf pulses: In black after a 1 second pulse with a strength of 3 $\upmu$T, in red an infinitly long pulse with the same strength, and in blue a 1 second pulse with a strength of 10 $\upmu$T. The bottom axis shows the detuning $\Delta \omega$, while the bottom axis shows the corresponding distance to the resonant slice, calculated using Eq. \ref{eq:dslice} assuming a magnetic field gradient $\nabla_r B_0 = 5\cdot 10^4$ T/m.}
\label{figure:Eq4p18}
\end{figure}

The consequences of Eq. \ref{eq:4p18} can be seen in Fig. \ref{figure:Eq4p18}. Depending on the precise pulse parameters, even the spins that do not meet the resonance condition by a detuning $\Delta \omega$ can lose (part of) their magnetization due to the rf pulse. The calculation is done assuming $T_1$ = 25 s and $T_2$ = 0.15 ms, typical values for copper at $T$ = 40 mK \cite{wagenaar2016}. The detuning can be translated to a distance to the resonant slice (the region where $\Delta \omega = 0$) using
\begin{equation} \label{eq:dslice}
d \approx \frac{\Delta \omega}{\gamma \nabla_r B_0}
\end{equation}
where $\nabla_r B_0$ is the gradient of the magnetic field in the radial direction. 

\subsection{Calculation of frequency shifts} \label{sec:freq_shift}
To calculate the frequency shift $\Delta f_0$ due to the saturation of the magnetization of the spins in resonance, we first look at the shift of the cantilever resonance frequency due to the coupling with a single spin. For this we follow a recent theoretical analysis of the magnetic coupling between a paramagnetic spin and the cantilever by \citeauthor{voogd2017}. In our case, where the frequency of the rf pulse $\omega_{rf} \gg \frac{1}{T_2}$ and $\omega T_1 \gg 1$, a single spin induces a stiffness shift given by
\begin{equation} \label{eq:Dk}
\Delta k = \langle m \rangle \left( |\bm{B''_{||B_0}}| + \frac{1}{B_0}|\bm{B'_{\perp B_0}}|^2 \right)
\end{equation}
The primes and double primes refer to the first and second derivative, respectively, with respect to the fundamental direction of motion of the cantilever. $|\bm{B''_{||B_0}}|$ is the component along $B_0$.  $|\bm{B'_{\perp B_0}}|$ is the perpendicular component. $\langle m \rangle$ is the mean Boltzmann polarization.

The effect of a rf pulse is to partially remove the magnetization of the spins by an amount given by:
\begin{align}
	\Delta m &= \langle m \rangle - m_z \\
	&= \langle m \rangle \left(1 - p_z \right) \left( 1 - e^{-\frac{t_p}{T_1 p_z}} \right),
\end{align}
where we set $m_0$ equal to $\langle m \rangle$, i.e. we assume the system is in thermal equillibrium before the pulse such that the initial magnetization is equal to the Boltzmann polarization. Please be reminded that $\Delta m$ is position dependent via $p_z$ due to the detuning $\Delta \omega$, which increases with the distance to the resonant slice and also depends on the precise rf pulse parameters. We can calculate the total measured frequency shift after a rf pulse by integrating over all spins in the sample including the position dependent demagnetization $\Delta m$:
\begin{equation} \label{eq:Df0}
\Delta f_0 = - \frac{1}{2} \frac{f_0}{k_0} \rho \int \Delta m \left( |\bm{B''_{||B_0}}| + \frac{1}{B_0}|\bm{B'_{\perp B_0}}|^2 \right) dV,
\end{equation}
with $\rho$ = 85 spins/nm$^3$ the spin density of copper. Alternatively, one can also sum the contribution of individual voxels, as long as the size of the voxels is small compared to the effective resonant slice width.

\section{Frequency shifts measured in copper}
In this section, we present measured frequency shifts using the higher modes of our cantilever as a source for the rf-field, on the one hand to demonstrate that the higher modes can indeed be used to perform full-fledged saturation experiments in MRFM, and on the other to give some experimental verifications of the theory presented in the previous section.

\begin{figure}
\centering
\includegraphics[width=\columnwidth]{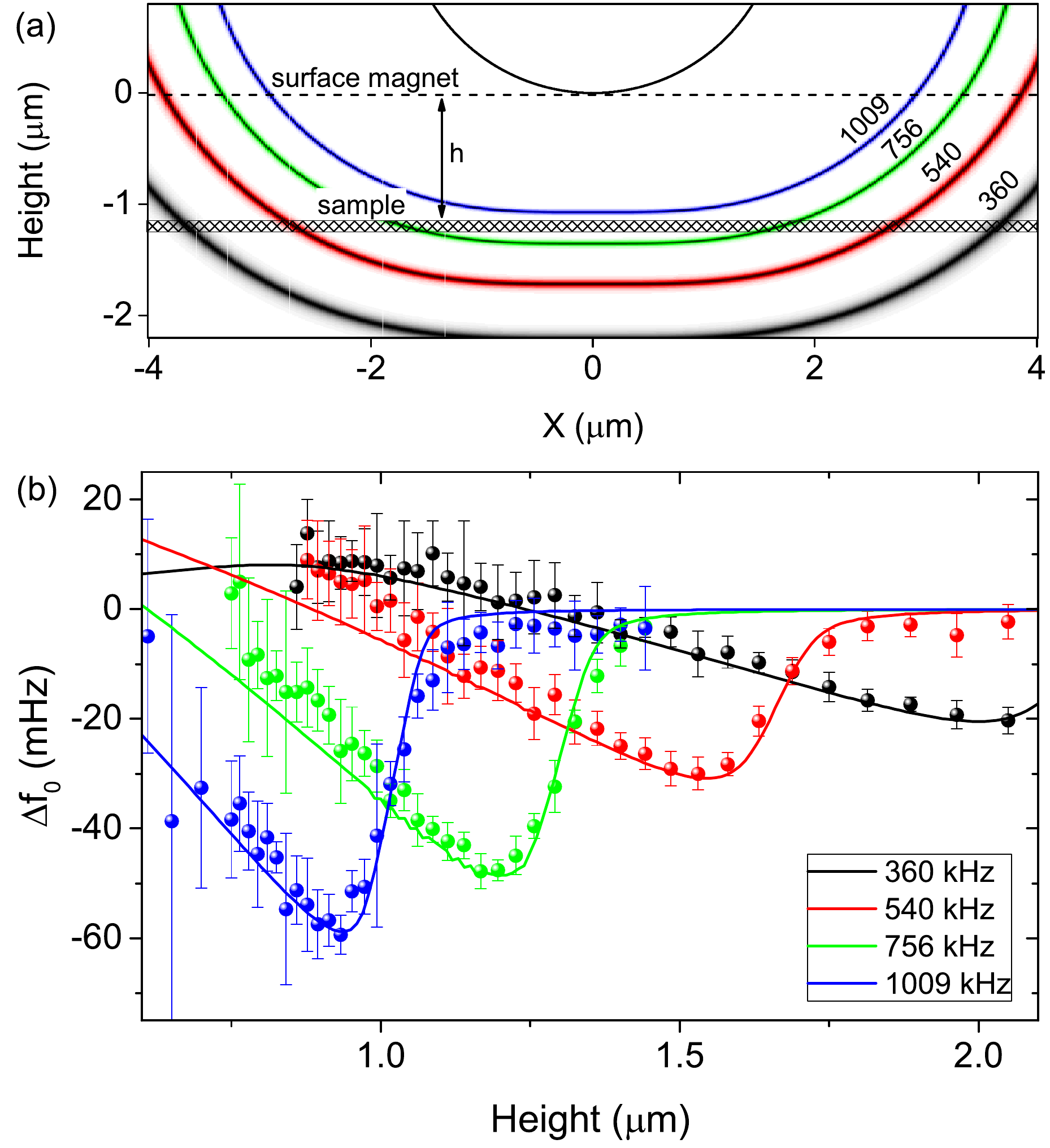}
\caption{(a) Positions of the resonant slices corresponding to the cantilever higher modes at 360 (black), 540 (red), 756 (green), and 1009 (blue) kHz. The black sphere at the top of the image  represents the cantilever magnet (radius 1.7 $\upmu$m, to scale). (b) Direct frequency shift $\Delta f_0$ versus height $h$ after exciting the spins by using the rf wire to drive the cantilever higher modes indicated in (a), measured at $T$ = 30 mK. Solid lines are the calculated signals for a pulse duration $t_p$ = 0.3 s, and $B_{rf}$ a free parameter. The error bars indicate the standard deviation of 10 single-shot measurements.}
\label{figure:Sig_vs_H}
\end{figure}

We demonstrate the effectiveness of using the higher modes of the cantilever as rf field source, by exciting 4 different higher modes of the cantilever by sending a current of 21 $\upmu$A\textsubscript{rms} through the rf wire. The frequencies of the selected higher modes are 360 kHz, 540 kHz, 756 kHz, and 1.009 MHz. The position of the resonant slices corresponding to these frequencies are shown in Fig. \ref{figure:Sig_vs_H}(a). The height of the magnet above the sample determines which of the resonant slices is in the sample, and how much signal each of these slices produces. In Fig. \ref{figure:Sig_vs_H}(b) we show the measured direct frequency shift $\Delta f_0$ as a function of the height for each of the higher modes, averaging over 10 single measurements. The error bars are determined by fitting 10 single-shot measurements and calculating the standard deviation of the fitted $\Delta f_0$. The solid lines in the figure are the calculated signals based on Eq. \ref{eq:Df0} using $t_p$ = 0.3 s. As the precise amplitude of the mechanically generated rf field is difficult to control since it depends on the distance between the magnet and the rf wire, the height of the magnet above the sample, and the Q-factor of the higher mode, the strength of the rf field is the only free fitting parameter. From the fits we obtain fields of 38, 35, 38, and 33 $\upmu$T for the 4 higher modes as mentioned before. Evidently, the different higher modes enter the sample at the predicted heights, with the correct overal magnitude of the direct frequency shift. The small deviation between the data and calculation at the lower heights probably results from a slightly changing $B_{rf}$. This measurement can be considered as a crude one-dimensional scan of the sample. Furthermore, considering that the current of 21 $\upmu$A\textsubscript{rms} corresponds to a field of only 0.2 $\upmu$T at the position of the cantilever, 7 $\upmu$m away from the rf wire, this measurement indicates that using the higher modes to generate the rf field results in an amplification of the rf field strength of more than a factor of 160. No heating was observed on the sample holder, indicating a dissipated power $<$ 1 nW.

\begin{figure}
\centering
\includegraphics[width=\columnwidth]{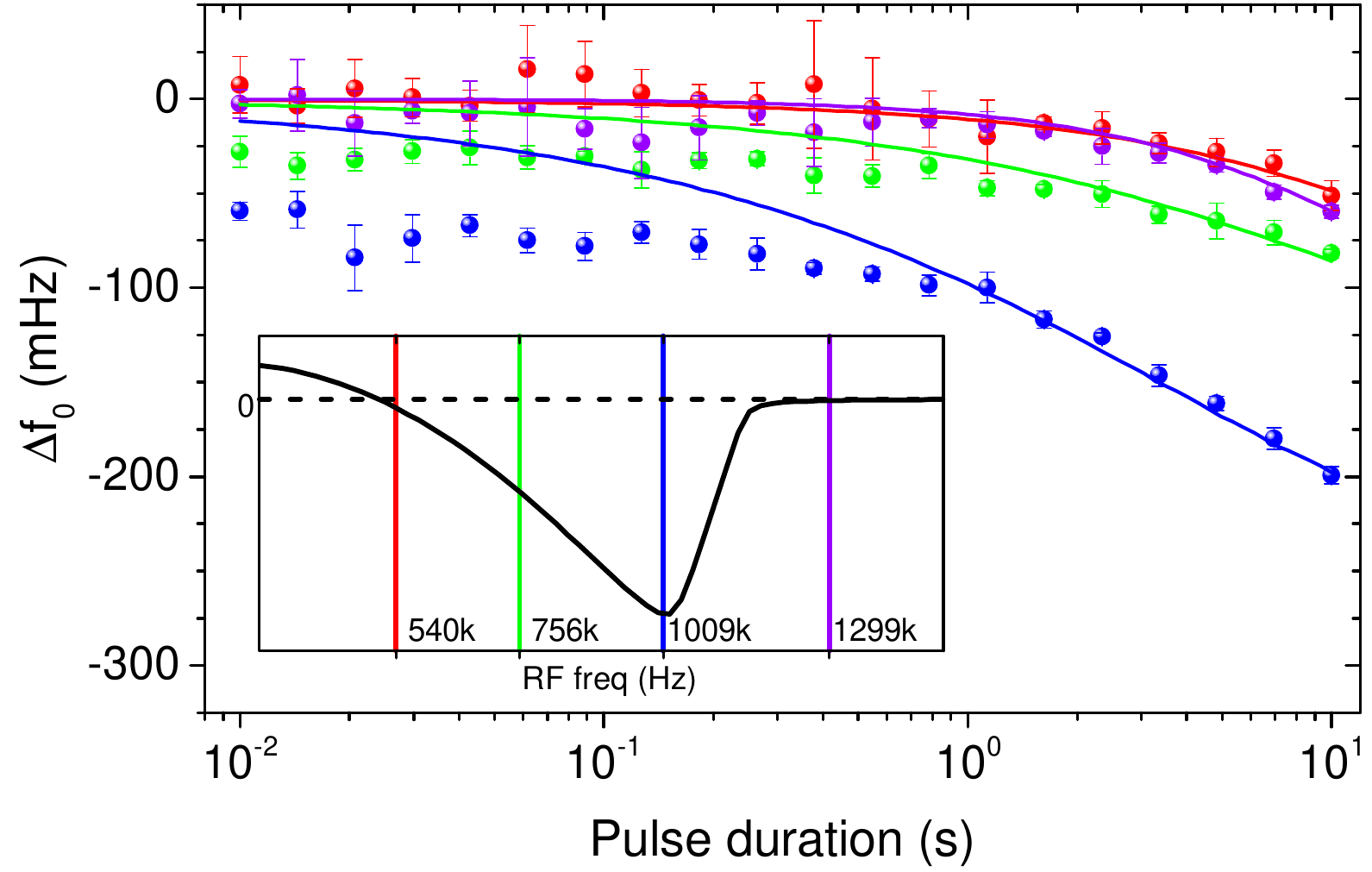}
\caption{Effect of the excitation pulse duration $t_p$ on the measured direct frequency shift $\Delta f_0$ for the cantilever higher modes at 540 (red), 756 (green), 1009 (blue), and 1299 (purple) kHz, measured at $h$ = 0.95 $\upmu$m and $T$ = 30 mK. The inset shows the calculated direct frequency shift as a function of the rf frequency, and also shows the position of the higher modes in this calculation. As $t_p$ increases, the resonant slice broadens and the direct frequency shift increases as expected from the resonant slice positions indicated in the inset. The error bars indicate the standard deviation of 5 single-shot measurements.}
\label{figure:Cu_HM-Sig-vs-Tp}
\end{figure}

We can further demonstrate the effect of the pulse parameters on the effective resonance slice width by doing a variation on the previous experiment. We now keep the sample at a constant height, and vary the duration of the rf current used to excite each of the higher modes in order to broaden the resonant slice. By comparing the measured increase of the signal for the various higher modes to the signal we expect from Eqs. \ref{eq:4p18} and \ref{eq:Df0}, we can confirm the applicability of these equations. This experiment is shown in Fig. \ref{figure:Cu_HM-Sig-vs-Tp}. The inset shows the calculated frequency shift as a function of the rf frequency, as well as the position of the higher modes. From the inset we see that for short pulses (a narrow resonant slice) we expect no signal from the 540 kHz and 1.299 MHz higher modes, some signal from the 756 kHz higher mode, and most signal from the 1.009 MHz higher mode. This behaviour is also observed in the main figure, where the solid lines are the calculated frequency shifts based on Eq. \ref{eq:Df0}. As $t_p$ is increased, even the resonant slices whose center is not in the sample broaden enough that off-resonant spins start to create measurable frequency shifts, with a good correlation between theory and experiment. The mismatch between the measured and calculated signal for very short pulse durations is attributed to the large Q-factor of the higher modes, which can be as high as $10^6$, resulting in characteristic time constants of up to 1 second. In that case, driving the higher mode for a very short time still results in a long effective pulse duration determined by the slow ringdown of the higher mode.

\section{Demonstration of volume sensitivity}

\begin{figure}
\centering
\includegraphics[width=\columnwidth]{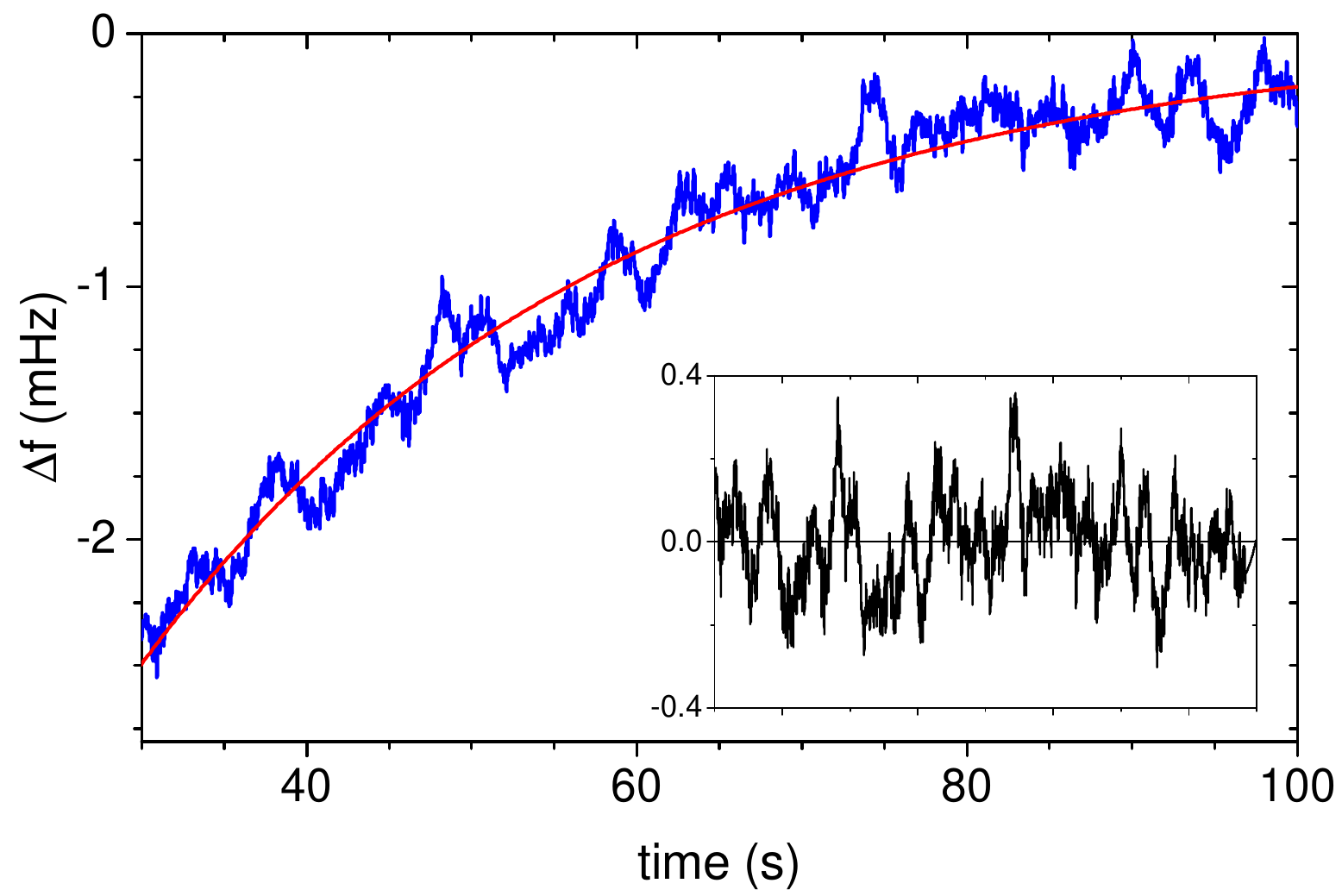}
\caption{Measured relaxation curve (1 Hz low-pass filter, 410 averages) measured at $h$ = 1.0 $\upmu$m and $T$ = 21 mK, for a pulse at frequency 882 kHz with $B_{rf}$ = 172 $\upmu$T and $t_p$ = 80 $\upmu$s. The solid red line is a fit to Eq. \ref{eq:exponential}, from which we extract $\Delta f_0$ = -5.4 mHz. The inset shows the difference between the data and the exponential fit, indicating a standard deviation of the measured frequency shift of 0.1 mHz.}
\label{figure:Cu_Ultimate_Res}
\end{figure}

As shown in Fig. \ref{figure:PLL_Noise}, we have a very clean frequency noise spectrum. To make full use of this, we have attempted to determine our optimal frequency resolution. To achieve this, we make a small adjustment to the measurement scheme, by switching off the cantilever drive a couple of seconds before we apply the rf pulse. The amplitude of the fundamental mode decays quickly due to the relatively low Q-factor of the fundamental mode close to the sample. By the time the pulse is sent, the amplitude of the cantilever is thermally limited to less than 0.1 nm. Directly after the pulse, the cantilever drive is switched back on to measure the resonance frequency shift. In this way, we prevent broadening of the resonance slice due to the cantilever amplitude of about 30 nm\textsubscript{rms}, and are able to achieve very narrow resonance slices. Fig. \ref{figure:Cu_Ultimate_Res} shows the relaxation curve measured at $T$ = 21 mK and $h$ = 1.0 $\upmu$m, after an 882 kHz rf pulse with $B_{rf}$ = 172 $\upmu$T and $t_p$ = 80 $\upmu$s. The blue curve shows the result of 410 averages with a total measurement time of over 10 hours, while the red curve is a fit to the data following Eq. \ref{eq:exponential}, from which we extract a direct frequency shift of -5.4 mHz. The inset shows the difference between the measured data and the fit, indicating that we can measure the frequency shift with a standard deviation of 0.1 mHz, consistent with the integrated frequency noise calculated from Fig. \ref{figure:PLL_Noise} and the number of averages. 

We can try to estimate the total detection volume that was necessary to generate this signal. In order to do so, we make the simplifying assumption that there exists a critical detuning $\Delta \omega_C$ such that all spins at a detuning smaller than the critical detuning (i.e. spins that feel a magnetic field between $B_0 - \Delta \omega_C / \gamma$ and $B_0 + \Delta \omega_C / \gamma$) are fully saturated, and spins at a detuning larger than the critical detuning are completely unaffected by the pulse. We then calculate the signal for various values of $\Delta \omega_C$ until we find the value for which the calculation matches the experiment. By dividing the sample in small voxels and summing all voxels that satisfy the condition specified above for the correct $\Delta \omega_C$, we find an estimate for the detection volume.

For the data presented in Fig. \ref{figure:Cu_Ultimate_Res} we find that this signal is the result of a critical detuning $\Delta \omega_C / (2\pi)$ = 2.1 kHz, equivalent to a resonant slice with a full width of approximately 4 nm. This corresponds to a total detection volume of (152 nm)$^3$, with a noise floor equal to (40 nm)$^3$. This volume contains a total of $5.6 \cdot 10^6$ spins at a Boltzmann polarization of about 0.3 \%, corresponding to about $1.6 \cdot 10^4$ fully polarized copper nuclear spins.

\section{Imaging protons}

With the volume sensitivities achieved on copper as demonstrated in the previous section, it is worthwhile to discuss what such an experiment would look like for a sample containing protons, the prime target spin for imaging purposes. Therefore, in this section we will calculate the signals that can be expected from a proton-rich sample, under the assumption that it is possible to achieve the same low frequency noise as in the current experiment on copper. $^1$H spins have spin $S$ = 1/2, gyromagnetic ratio $\gamma_H / (2\pi) = 42.6$ MHz/T, and a magnetic moment $\mu_H = 1.41 \cdot 10^{-26} J/T$. For MRFM, proton spins are generally a bit more favourable than copper spins, as the higher gyromagnetic ratio and magnetic moment mean a higher Boltzmann polarization and a larger coupling between a single spin and the cantilever. We assume a proton spin density $\rho_H$ = 50 spins/nm$^3$, a typical value for biological tissue and polymers \cite{degen2009,isaac2016}. Furthermore, we assume $T_1$ = 30 s and $T_2$ = 0.1 ms. Note that the exact values for the relaxation times do not matter that much as long as the conditions used for the derivation of Eqs. \ref{eq:4p18} and \ref{eq:Dk} are met, and the rf pulse duration is short compared to $T_1$.

We calculate the total volume necessary to get a frequency shift of 1.8 mHz, a signal that can be measured in a single shot experiment assuming the SNR achieved on the copper, and 0.5 mHz, which can be measured within 30 minutes ($\sim$ 15 averages). The results can be found in Table \ref{table:protons}. We considered three different experimental configurations, where we vary the size of the magnet in order to increase the field gradients and thereby the signal per spin. The first configuration is a replication of the experimental parameters as used for the copper measurement from Fig. \ref{figure:Cu_Ultimate_Res}: A saturation experiment performed at a height of 1.0 $\upmu$m and a temperature of 21 mK. The optimal signal at this height is found for a rf frequency of 3.5 MHz (about a factor of 4 higher than the rf frequency used for the copper due to the higher gyromagnetic ratio). The other two configuration are simulations with magnets with radii of 1.0 $\upmu$m and 0.5 $\upmu$m. To make a fair comparison, we calculate the signal for the same Larmor frequency of 3.5 MHz, which dictates measurement heights of 0.56 $\upmu$m and 0.24 $\upmu$m. All unmentioned parameters are kept constant. The predicted detection volumes for the different configuration are shown in Table. \ref{table:protons}.

\begin{table}
		\begin{tabular}{c c c c c}
		\hline\hline
		$R_0$	& $h$	& $\nabla_r B_0$	& $V_{ss}$	& $V_{30min}$ \\	 
		\hline
		1.7 $\upmu$m ~ 	& ~ 1.00 $\upmu$m ~ 	& 	100 $\upmu$T/nm  	&	~ (84 nm)$^3$ ~ 	& ~ (55 nm)$^3$ \\
		1.0 $\upmu$m ~ 	& ~ 0.56 $\upmu$m ~  	& 	170 $\upmu$T/nm  	&	~ (59 nm)$^3$ ~ 	& ~ (39 nm)$^3$ \\
		0.5 $\upmu$m ~ 	& ~ 0.24 $\upmu$m ~  	& 	370 $\upmu$T/nm  	&	~ (39 nm)$^3$ ~ 	& ~ (25 nm)$^3$ \\
		\hline\hline 
		\end{tabular}
	\caption{Calculated volume sensities $V_{ss}$ (volume required for 1.8 mHz frequency shift) and $V_{30min,DNP}$ (volume required for a 0.5 mHz frequency shift). Calculations are done for sample temperature $T$ = 21 mK and rf frequency $\omega_{rf}/(2 \pi)$ = 3.5 MHz. The radial magnetic field gradient $\nabla_r B_0$ is calculated at 50 nm below the surface of the sample.} 
	\label{table:protons}
\end{table}

Clearly, decreasing the size of the magnetic particle will enhance the volume sensitivity, but there is a fundamental limit: the experiment described here relies on removing the Boltzmann polarization of the sample, but as the detection volume goes down, we enter the regime where statistical polarization becomes dominant. The critical volume $V_c$ for this transition is given by \cite{herzog2014}
\begin{equation}
V_c = \frac{4}{\rho_H} \left( \frac{k_B T}{\hbar \gamma B_0} \right)^2,
\end{equation}
where it is assumed that the thermal energy is much larger than the Zeeman splitting. For a temperature of 21 mK and a Larmor frequency of 3.5 MHz, $V_c \sim$ (11 nm)$^3$. Below this detection volume, measurements of the direct frequency shift would average to zero.

However, further enhancement of the volume sensitivity can still be achieved by increasing the Boltzmann polarization of the protons. This can be done by working at higher Larmor frequencies by decreasing the tip-sample separation, or by applying a strong external magnetic field. An external magnetic field of 8T would increase the Boltzmann polarization by roughly a factor of 100, but applying external magnetic fields in combination with our SQUID-based detection is challenging due to our extreme sensitivity to magnetic noise. An appealing alternative is to use dynamical nuclear polarization (DNP), as was recently demonstrated for MRFM by \citeauthor{isaac2016}. For suitable samples, e.g. nitroxide-doped polystyrene, DNP can be used to transfer polarization from electron spins to nuclei. The maximum enhancement of the nuclear polarization that can be achieved using this mechanism is given by $\epsilon = \gamma_e / \gamma_H = 660$. However, for protons at a Larmor frequency of 3.5 MHz and temperature of 21 mK the initial Boltzmann polarization is about 0.4\%, so our maximal enhancement is limited to a factor 250. Table. \ref{table:DNP} shows the calculated volume sensitivities if we are able to use DNP to enhance the nuclear polarization, for the cases where we achieve DNP efficiencies of 10\% and 100\%. Even for the more realistic assumption of 10\% efficiency, we find that a volume sensitivity below (10 nm)$^3$ could be possible. This voxel size would make imaging based on measurements of the Boltzmann polarization a viable approach to image biological samples, without the demand for high rf field amplitudes and continuous application of this field, as was the case for previous amplitude-based imaging \cite{degen2009}.

\begin{table}
		\begin{tabular}{c c c c c}
		\hline\hline
		$R_0$	& 	$h$	&  	$DNP_{eff}$	& $V_{ss,DNP}$	& 	$V_{30min,DNP}$ \\	 
		\hline
		~ 1.7 $\upmu$m ~ 	& ~ 1.00 $\upmu$m ~   	& ~ 10\% ~  	& ~ (21 nm)$^3$ ~	& ~ (14 nm)$^3$ ~ \\
		~ 1.7 $\upmu$m ~ 	& ~ 1.00 $\upmu$m ~   	& ~ 100\% ~  	& ~ (13 nm)$^3$ ~	& ~ (8.7 nm)$^3$ ~ \\
		~ 1.0 $\upmu$m ~  	& ~ 0.56 $\upmu$m ~   	& ~ 10\% ~   	& ~ (15 nm)$^3$ ~	& ~ (10 nm)$^3$ ~ \\
		~ 1.0 $\upmu$m ~  	& ~ 0.56 $\upmu$m ~   	& ~ 100\% ~   	& ~ (9.4 nm)$^3$ ~	& ~ (6.1 nm)$^3$ ~ \\
		~ 0.5 $\upmu$m ~  	& ~ 0.24 $\upmu$m ~   	& ~ 10\% ~   	& ~ (9.6 nm)$^3$ ~	& ~ (6.2 nm)$^3$ ~ \\
		~ 0.5 $\upmu$m ~  	& ~ 0.24 $\upmu$m ~   	& ~ 100\% ~   	& ~ (6.1 nm)$^3$ ~	& ~ (4.0 nm)$^3$ ~ \\
		\hline\hline 
		\end{tabular}
	\caption{Calculated volume sensities $V_{ss,DNP}$ and $V_{30min,DNP}$ including DNP to enhance the nuclear polarization with an efficiency $DNP_{eff}$. Calculations are done for sample temperature $T$ = 21 mK and rf frequency $\omega_{rf}/(2 \pi)$ = 3.5 MHz.} 
	\label{table:DNP}
\end{table}

Of course, there are some potential pitfalls that should be considered. First of all, we have assumed that the frequency noise spectrum shown in Fig. \ref{figure:PLL_Noise} can be maintained. However, large 1/f noise has been reported at 4K on insulating samples like polymers, attributed to dielectric fluctuations \cite{yazdanian2008,hoepker2011}. This frequency noise scales with the square of the charge difference between the sample and the tip. Therefore, we believe it can be avoided, either by properly grounding both the tip and sample, but also by biasing the tip to tune away any charge difference \cite{yazdanian2009,moore2009}.

A second limitation is that for the current experiment we require $T_1$ times to be between several seconds and minutes. When $T_1$ is shorten than several seconds, it becomes comparable to other time constants in our setup (e.g. the thermal time constant of the sample holder), making it difficult to extract the signal. When $T_1$ becomes longer than minutes, averaging measurements to increase the SNR will become very time-consuming, although the total measurement time may come down by using multiple resonance slices \cite{oosterkamp2010,moores2015}. Plus, as the duration of a measurement increases, 1/f noise will increasingly become a limiting factor. $T_1$ times within the desired range for suitable proton samples are reported at low temperatures \cite{chen2013,isaac2016}. For very pure samples with long $T_1$ times, appropriate doping of the sample with impurities can be used to reduce the relaxation time \cite{grapengeter1980}.

\begin{figure}
\centering
\includegraphics[width=\columnwidth]{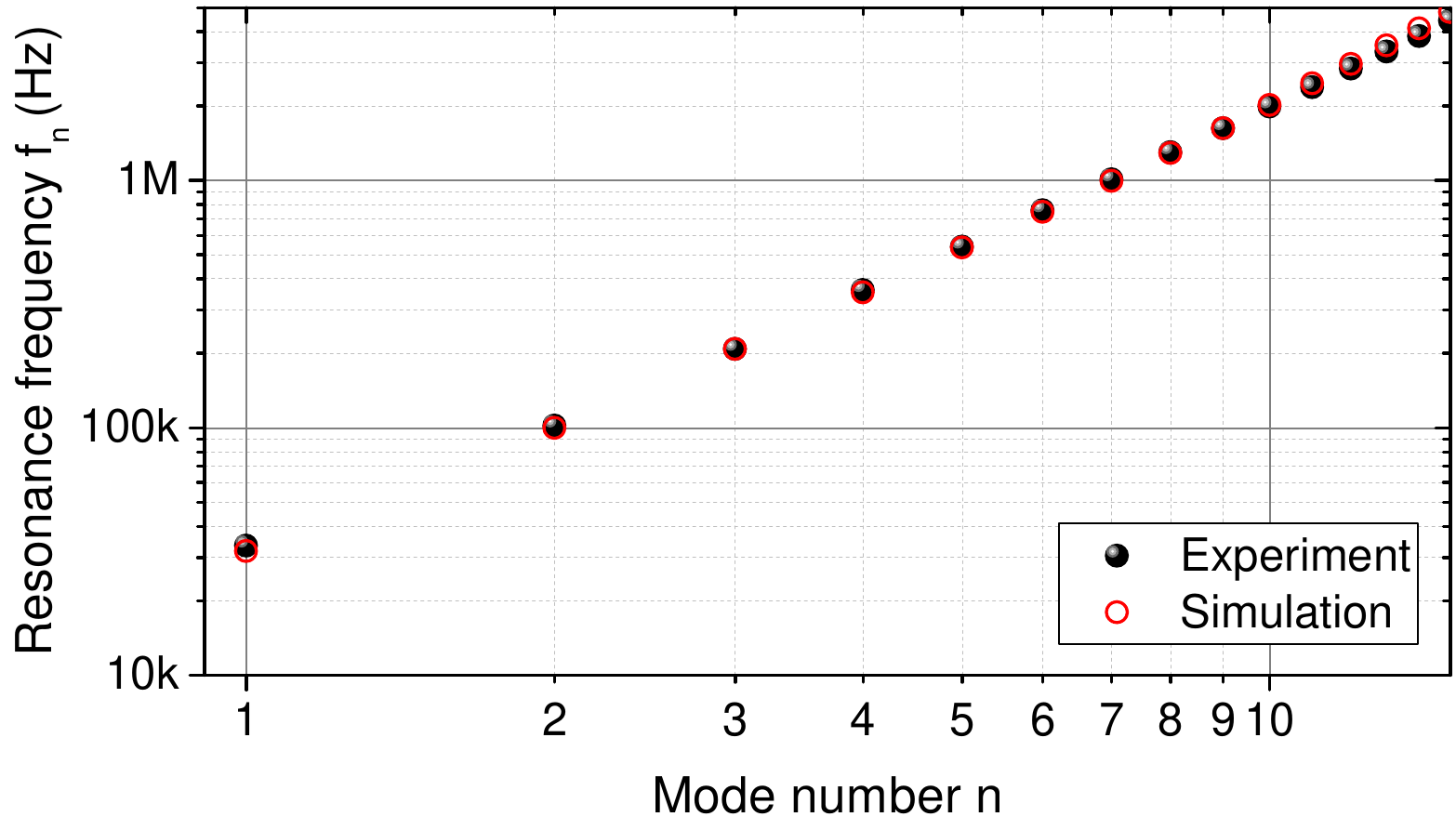}
\caption{Measured higher mode resonance frequencies of the cantilever, together with the mode frequenties obtained from finite element calculations. The highest resonance mode investigated is the 15\textsuperscript{th} mode located at $f_{15} = 4.4$ MHz. In the simulations we only consider higher modes that vibrate in the soft direction of the cantilever.}
\label{figure:Higher_Modes}
\end{figure}

The final challenge is to maintain the low operating temperatures required for the low frequency noise floor while sending rf pulses in the MHz range. The power dissipated by the rf pulse, even when using a superconducting rf wire, increases with the frequency. To apply a 0.1 mT rf pulse at a sample located 5 $\upmu$m from the rf wire at 3.5 MHz, we measure a dissipation of approximately 3 $\upmu$W in our setup. A continuous power pulse with this level of dissipation would locally heat the sample to over 100 mK. We can avoid this source of dissipation by using the higher modes of the cantilever, which can be excited up to the 15\textsuperscript{th} mode at 4.4 MHz and possibly beyond. In Fig. \ref{figure:Higher_Modes} we show the frequencies of the higher flexural modes together with the calculated frequencies obtained from finite element calculations. The estimated dissipation from the motion of a higher mode is well below 1 fW, since we measure the higher modes to have Q-factors approaching one million \cite{wagenaar2017}.

\section{Conclusions}
We have used the time-dependent solution of the Bloch equation to derive a consise equation to calculate the frequency shifts in MRFM experiments, and applied this to saturation experiments on a thin copper film. By using the higher modes of the cantilever as a source for the rf fields, we have demonstrated that it is possible to make one-dimensional scans of the copper film with near negligible dissipation, and that the measured direct frequency shifts are well reproduced by the presented theory. Finally, we have shown that we have measured a frequency-shift signal with a volume sensitivity of (40 nm)$^3$. We have done all this at temperatures as low as 21 mK, made possible by the SQUID-based detection of the cantilever and the low power saturation protocol in combination with the mechanical generation of the rf fields.

The achieved volume sensitivity opens up the way for imaging based on measurements of the Boltzmann polarization, which could allow for high resolution imaging due to the direct gain from lower temperatures, and the favourable averaging compared to statistical polarization based imaging. We have shown that modest technical changes to our current setup can allow for experiments on protons with a spatial resolution of (25 nm)$^3$, and that increasing the polarization, for instance using DNP, can improve the resolution even further to below (10 nm)$^3$. The magnet-on-tip geometry allows for a larger choice in available samples, as it is still an open question whether interesting biological samples can be attached to an ultrasoft MRFM cantilever for approaches using the sample-on-tip geometry. When it is possible to measure on biological samples with the same low frequency noise as achieved in the current experiment, high-resolution Boltzmann polarization based magnetic resonance imaging at milliKelvin temperatures in a magnet-on-tip geometry could become a reality. 

\section{acknowledgement}
The authors thank K. Heeck, M. Camp, G. Koning, F. Schenkel, D.~J. van der Zalm, J.~P. Koning, and L. Crama for technical support. The authors thank D.~J. Thoen, T.~M. Klapwijk, and A. Endo for providing us with the NbTiN. The authors thank T.~H.~A. van der Reep for valuable discussions and proofreading the manuscript. This work is supported by the Netherlands Organisation for Scientific Research (NWO) through a VICI fellowship to T.~H.~O., and through the Nanofront program.

\bibliography{NMR_Copper_BIB}

\begin{thebibliography}{39}%
\makeatletter
\providecommand \@ifxundefined [1]{%
 \@ifx{#1\undefined}
}%
\providecommand \@ifnum [1]{%
 \ifnum #1\expandafter \@firstoftwo
 \else \expandafter \@secondoftwo
 \fi
}%
\providecommand \@ifx [1]{%
 \ifx #1\expandafter \@firstoftwo
 \else \expandafter \@secondoftwo
 \fi
}%
\providecommand \natexlab [1]{#1}%
\providecommand \enquote  [1]{``#1''}%
\providecommand \bibnamefont  [1]{#1}%
\providecommand \bibfnamefont [1]{#1}%
\providecommand \citenamefont [1]{#1}%
\providecommand \href@noop [0]{\@secondoftwo}%
\providecommand \href [0]{\begingroup \@sanitize@url \@href}%
\providecommand \@href[1]{\@@startlink{#1}\@@href}%
\providecommand \@@href[1]{\endgroup#1\@@endlink}%
\providecommand \@sanitize@url [0]{\catcode `\\12\catcode `\$12\catcode
  `\&12\catcode `\#12\catcode `\^12\catcode `\_12\catcode `\%12\relax}%
\providecommand \@@startlink[1]{}%
\providecommand \@@endlink[0]{}%
\providecommand \url  [0]{\begingroup\@sanitize@url \@url }%
\providecommand \@url [1]{\endgroup\@href {#1}{\urlprefix }}%
\providecommand \urlprefix  [0]{URL }%
\providecommand \Eprint [0]{\href }%
\providecommand \doibase [0]{http://dx.doi.org/}%
\providecommand \selectlanguage [0]{\@gobble}%
\providecommand \bibinfo  [0]{\@secondoftwo}%
\providecommand \bibfield  [0]{\@secondoftwo}%
\providecommand \translation [1]{[#1]}%
\providecommand \BibitemOpen [0]{}%
\providecommand \bibitemStop [0]{}%
\providecommand \bibitemNoStop [0]{.\EOS\space}%
\providecommand \EOS [0]{\spacefactor3000\relax}%
\providecommand \BibitemShut  [1]{\csname bibitem#1\endcsname}%
\let\auto@bib@innerbib\@empty
\bibitem [{\citenamefont {Z{\"u}ger}\ and\ \citenamefont
  {Rugar}(1993)}]{zuger1993}%
  \BibitemOpen
  \bibfield  {author} {\bibinfo {author} {\bibfnamefont {O.}~\bibnamefont
  {Z{\"u}ger}}\ and\ \bibinfo {author} {\bibfnamefont {D.}~\bibnamefont
  {Rugar}},\ }\href {\doibase 10.1063/1.110460} {\bibfield  {journal} {\bibinfo
   {journal} {Applied Physics Letters}\ }\textbf {\bibinfo {volume} {63}},\
  \bibinfo {pages} {2496} (\bibinfo {year} {1993})}\BibitemShut {NoStop}%
\bibitem [{\citenamefont {Rugar}\ \emph {et~al.}(1994)\citenamefont {Rugar},
  \citenamefont {Z{\"u}ger}, \citenamefont {Hoen}, \citenamefont {Yannoni},
  \citenamefont {Vieth},\ and\ \citenamefont {Kendrick}}]{rugar1994}%
  \BibitemOpen
  \bibfield  {author} {\bibinfo {author} {\bibfnamefont {D.}~\bibnamefont
  {Rugar}}, \bibinfo {author} {\bibfnamefont {O.}~\bibnamefont {Z{\"u}ger}},
  \bibinfo {author} {\bibfnamefont {S.}~\bibnamefont {Hoen}}, \bibinfo {author}
  {\bibfnamefont {C.~S.}\ \bibnamefont {Yannoni}}, \bibinfo {author}
  {\bibfnamefont {H.}~\bibnamefont {Vieth}}, \ and\ \bibinfo {author}
  {\bibfnamefont {R.~D.}\ \bibnamefont {Kendrick}},\ }\href {\doibase
  10.1126/science.264.5165.1560} {\bibfield  {journal} {\bibinfo  {journal}
  {Science}\ }\textbf {\bibinfo {volume} {264}},\ \bibinfo {pages} {1560}
  (\bibinfo {year} {1994})}\BibitemShut {NoStop}%
\bibitem [{\citenamefont {Sidles}\ \emph {et~al.}(1995)\citenamefont {Sidles},
  \citenamefont {Garbini}, \citenamefont {Bruland}, \citenamefont {Rugar},
  \citenamefont {Z{\"u}ger}, \citenamefont {Hoen},\ and\ \citenamefont
  {Yannoni}}]{sidles1995}%
  \BibitemOpen
  \bibfield  {author} {\bibinfo {author} {\bibfnamefont {J.~A.}\ \bibnamefont
  {Sidles}}, \bibinfo {author} {\bibfnamefont {J.~L.}\ \bibnamefont {Garbini}},
  \bibinfo {author} {\bibfnamefont {K.~J.}\ \bibnamefont {Bruland}}, \bibinfo
  {author} {\bibfnamefont {D.}~\bibnamefont {Rugar}}, \bibinfo {author}
  {\bibfnamefont {O.}~\bibnamefont {Z{\"u}ger}}, \bibinfo {author}
  {\bibfnamefont {S.}~\bibnamefont {Hoen}}, \ and\ \bibinfo {author}
  {\bibfnamefont {C.~S.}\ \bibnamefont {Yannoni}},\ }\href {\doibase
  10.1103/RevModPhys.67.249} {\bibfield  {journal} {\bibinfo  {journal}
  {Reviews of Modern Physics}\ }\textbf {\bibinfo {volume} {67}},\ \bibinfo
  {pages} {249} (\bibinfo {year} {1995})}\BibitemShut {NoStop}%
\bibitem [{\citenamefont {Rugar}\ \emph {et~al.}(2004)\citenamefont {Rugar},
  \citenamefont {Budakian}, \citenamefont {Mamin},\ and\ \citenamefont
  {Chui}}]{rugar2004}%
  \BibitemOpen
  \bibfield  {author} {\bibinfo {author} {\bibfnamefont {D.}~\bibnamefont
  {Rugar}}, \bibinfo {author} {\bibfnamefont {R.}~\bibnamefont {Budakian}},
  \bibinfo {author} {\bibfnamefont {H.~J.}\ \bibnamefont {Mamin}}, \ and\
  \bibinfo {author} {\bibfnamefont {B.~W.}\ \bibnamefont {Chui}},\ }\href
  {\doibase 10.1038/nature02658} {\bibfield  {journal} {\bibinfo  {journal}
  {Nature}\ }\textbf {\bibinfo {volume} {430}},\ \bibinfo {pages} {329}
  (\bibinfo {year} {2004})}\BibitemShut {NoStop}%
\bibitem [{\citenamefont {Degen}\ \emph {et~al.}(2009)\citenamefont {Degen},
  \citenamefont {Poggio}, \citenamefont {Mamin}, \citenamefont {Rettner},\ and\
  \citenamefont {Rugar}}]{degen2009}%
  \BibitemOpen
  \bibfield  {author} {\bibinfo {author} {\bibfnamefont {C.~L.}\ \bibnamefont
  {Degen}}, \bibinfo {author} {\bibfnamefont {M.}~\bibnamefont {Poggio}},
  \bibinfo {author} {\bibfnamefont {H.~J.}\ \bibnamefont {Mamin}}, \bibinfo
  {author} {\bibfnamefont {C.~T.}\ \bibnamefont {Rettner}}, \ and\ \bibinfo
  {author} {\bibfnamefont {D.}~\bibnamefont {Rugar}},\ }\href {\doibase
  10.1073/pnas.0812068106} {\bibfield  {journal} {\bibinfo  {journal}
  {Proceedings of the National Academy of Sciences}\ }\textbf {\bibinfo
  {volume} {106}},\ \bibinfo {pages} {1313} (\bibinfo {year}
  {2009})}\BibitemShut {NoStop}%
\bibitem [{\citenamefont {Rose}\ \emph {et~al.}(2018)\citenamefont {Rose},
  \citenamefont {Haas}, \citenamefont {Chen}, \citenamefont {Jeon},
  \citenamefont {Lauhon}, \citenamefont {Cory},\ and\ \citenamefont
  {Budakian}}]{rose2018}%
  \BibitemOpen
  \bibfield  {author} {\bibinfo {author} {\bibfnamefont {W.}~\bibnamefont
  {Rose}}, \bibinfo {author} {\bibfnamefont {H.}~\bibnamefont {Haas}}, \bibinfo
  {author} {\bibfnamefont {A.~Q.}\ \bibnamefont {Chen}}, \bibinfo {author}
  {\bibfnamefont {N.}~\bibnamefont {Jeon}}, \bibinfo {author} {\bibfnamefont
  {L.~J.}\ \bibnamefont {Lauhon}}, \bibinfo {author} {\bibfnamefont {D.~G.}\
  \bibnamefont {Cory}}, \ and\ \bibinfo {author} {\bibfnamefont
  {R.}~\bibnamefont {Budakian}},\ }\href {\doibase 10.1103/PhysRevX.8.011030}
  {\bibfield  {journal} {\bibinfo  {journal} {Physical Review X}\ }\textbf
  {\bibinfo {volume} {8}},\ \bibinfo {pages} {011030} (\bibinfo {year}
  {2018})}\BibitemShut {NoStop}%
\bibitem [{\citenamefont {Poggio}\ \emph {et~al.}(2007)\citenamefont {Poggio},
  \citenamefont {Degen}, \citenamefont {Rettner}, \citenamefont {Mamin},\ and\
  \citenamefont {Rugar}}]{poggio2007}%
  \BibitemOpen
  \bibfield  {author} {\bibinfo {author} {\bibfnamefont {M.}~\bibnamefont
  {Poggio}}, \bibinfo {author} {\bibfnamefont {C.~L.}\ \bibnamefont {Degen}},
  \bibinfo {author} {\bibfnamefont {C.~T.}\ \bibnamefont {Rettner}}, \bibinfo
  {author} {\bibfnamefont {H.~J.}\ \bibnamefont {Mamin}}, \ and\ \bibinfo
  {author} {\bibfnamefont {D.}~\bibnamefont {Rugar}},\ }\href {\doibase
  10.1063/1.2752536} {\bibfield  {journal} {\bibinfo  {journal} {Applied
  physics letters}\ }\textbf {\bibinfo {volume} {90}},\ \bibinfo {pages}
  {263111} (\bibinfo {year} {2007})}\BibitemShut {NoStop}%
\bibitem [{\citenamefont {Nichol}\ \emph {et~al.}(2012)\citenamefont {Nichol},
  \citenamefont {Hemesath}, \citenamefont {Lauhon},\ and\ \citenamefont
  {Budakian}}]{nichol2012}%
  \BibitemOpen
  \bibfield  {author} {\bibinfo {author} {\bibfnamefont {J.~M.}\ \bibnamefont
  {Nichol}}, \bibinfo {author} {\bibfnamefont {E.~R.}\ \bibnamefont
  {Hemesath}}, \bibinfo {author} {\bibfnamefont {L.~J.}\ \bibnamefont
  {Lauhon}}, \ and\ \bibinfo {author} {\bibfnamefont {R.}~\bibnamefont
  {Budakian}},\ }\href {\doibase 10.1103/PhysRevB.85.054414} {\bibfield
  {journal} {\bibinfo  {journal} {Physical Review B}\ }\textbf {\bibinfo
  {volume} {85}},\ \bibinfo {pages} {054414} (\bibinfo {year}
  {2012})}\BibitemShut {NoStop}%
\bibitem [{\citenamefont {Garner}\ \emph {et~al.}(2004)\citenamefont {Garner},
  \citenamefont {Kuehn}, \citenamefont {Dawlaty}, \citenamefont {Jenkins},\
  and\ \citenamefont {Marohn}}]{garner2004}%
  \BibitemOpen
  \bibfield  {author} {\bibinfo {author} {\bibfnamefont {S.~R.}\ \bibnamefont
  {Garner}}, \bibinfo {author} {\bibfnamefont {S.}~\bibnamefont {Kuehn}},
  \bibinfo {author} {\bibfnamefont {J.~M.}\ \bibnamefont {Dawlaty}}, \bibinfo
  {author} {\bibfnamefont {N.~E.}\ \bibnamefont {Jenkins}}, \ and\ \bibinfo
  {author} {\bibfnamefont {J.~A.}\ \bibnamefont {Marohn}},\ }\href {\doibase
  10.1063/1.1762700} {\bibfield  {journal} {\bibinfo  {journal} {Applied
  physics letters}\ }\textbf {\bibinfo {volume} {84}},\ \bibinfo {pages} {5091}
  (\bibinfo {year} {2004})}\BibitemShut {NoStop}%
\bibitem [{\citenamefont {Mamin}\ \emph {et~al.}(2007)\citenamefont {Mamin},
  \citenamefont {Poggio}, \citenamefont {Degen},\ and\ \citenamefont
  {Rugar}}]{mamin2007}%
  \BibitemOpen
  \bibfield  {author} {\bibinfo {author} {\bibfnamefont {H.~J.}\ \bibnamefont
  {Mamin}}, \bibinfo {author} {\bibfnamefont {M.}~\bibnamefont {Poggio}},
  \bibinfo {author} {\bibfnamefont {C.~L.}\ \bibnamefont {Degen}}, \ and\
  \bibinfo {author} {\bibfnamefont {D.}~\bibnamefont {Rugar}},\ }\href
  {\doibase 10.1038/nnano.2007.105} {\bibfield  {journal} {\bibinfo  {journal}
  {Nature nanotechnology}\ }\textbf {\bibinfo {volume} {2}},\ \bibinfo {pages}
  {301} (\bibinfo {year} {2007})}\BibitemShut {NoStop}%
\bibitem [{\citenamefont {Alexson}\ \emph {et~al.}(2012)\citenamefont
  {Alexson}, \citenamefont {Hickman}, \citenamefont {Marohn},\ and\
  \citenamefont {Smith}}]{alexson2012}%
  \BibitemOpen
  \bibfield  {author} {\bibinfo {author} {\bibfnamefont {D.~A.}\ \bibnamefont
  {Alexson}}, \bibinfo {author} {\bibfnamefont {S.~A.}\ \bibnamefont
  {Hickman}}, \bibinfo {author} {\bibfnamefont {J.~A.}\ \bibnamefont {Marohn}},
  \ and\ \bibinfo {author} {\bibfnamefont {D.~D.}\ \bibnamefont {Smith}},\
  }\href {\doibase 10.1063/1.4730610} {\bibfield  {journal} {\bibinfo
  {journal} {Applied physics letters}\ }\textbf {\bibinfo {volume} {101}},\
  \bibinfo {pages} {022103} (\bibinfo {year} {2012})}\BibitemShut {NoStop}%
\bibitem [{\citenamefont {Wagenaar}\ \emph {et~al.}(2016)\citenamefont
  {Wagenaar}, \citenamefont {den Haan}, \citenamefont {de~Voogd}, \citenamefont
  {Bossoni}, \citenamefont {de~Jong}, \citenamefont {de~Wit}, \citenamefont
  {Bastiaans}, \citenamefont {Thoen}, \citenamefont {Endo}, \citenamefont
  {Klapwijk}, \citenamefont {Zaanen},\ and\ \citenamefont
  {Oosterkamp}}]{wagenaar2016}%
  \BibitemOpen
  \bibfield  {author} {\bibinfo {author} {\bibfnamefont {J.~J.~T.}\
  \bibnamefont {Wagenaar}}, \bibinfo {author} {\bibfnamefont {A.~M.~J.}\
  \bibnamefont {den Haan}}, \bibinfo {author} {\bibfnamefont {J.~M.}\
  \bibnamefont {de~Voogd}}, \bibinfo {author} {\bibfnamefont {L.}~\bibnamefont
  {Bossoni}}, \bibinfo {author} {\bibfnamefont {T.~A.}\ \bibnamefont
  {de~Jong}}, \bibinfo {author} {\bibfnamefont {M.}~\bibnamefont {de~Wit}},
  \bibinfo {author} {\bibfnamefont {K.~M.}\ \bibnamefont {Bastiaans}}, \bibinfo
  {author} {\bibfnamefont {D.~J.}\ \bibnamefont {Thoen}}, \bibinfo {author}
  {\bibfnamefont {A.}~\bibnamefont {Endo}}, \bibinfo {author} {\bibfnamefont
  {T.~M.}\ \bibnamefont {Klapwijk}}, \bibinfo {author} {\bibfnamefont
  {J.}~\bibnamefont {Zaanen}}, \ and\ \bibinfo {author} {\bibfnamefont {T.~H.}\
  \bibnamefont {Oosterkamp}},\ }\href {\doibase
  10.1103/PhysRevApplied.6.014007} {\bibfield  {journal} {\bibinfo  {journal}
  {Physical Review Applied}\ }\textbf {\bibinfo {volume} {6}},\ \bibinfo
  {pages} {014007} (\bibinfo {year} {2016})}\BibitemShut {NoStop}%
\bibitem [{\citenamefont {Wagenaar}\ \emph {et~al.}(2017)\citenamefont
  {Wagenaar}, \citenamefont {den Haan}, \citenamefont {Donkersloot},
  \citenamefont {Marsman}, \citenamefont {de~Wit}, \citenamefont {Bossoni},\
  and\ \citenamefont {Oosterkamp}}]{wagenaar2017}%
  \BibitemOpen
  \bibfield  {author} {\bibinfo {author} {\bibfnamefont {J.~J.~T.}\
  \bibnamefont {Wagenaar}}, \bibinfo {author} {\bibfnamefont {A.~M.~J.}\
  \bibnamefont {den Haan}}, \bibinfo {author} {\bibfnamefont {R.~J.}\
  \bibnamefont {Donkersloot}}, \bibinfo {author} {\bibfnamefont
  {F.}~\bibnamefont {Marsman}}, \bibinfo {author} {\bibfnamefont
  {M.}~\bibnamefont {de~Wit}}, \bibinfo {author} {\bibfnamefont
  {L.}~\bibnamefont {Bossoni}}, \ and\ \bibinfo {author} {\bibfnamefont
  {T.~H.}\ \bibnamefont {Oosterkamp}},\ }\href {\doibase
  10.1103/PhysRevApplied.7.024019} {\bibfield  {journal} {\bibinfo  {journal}
  {Physical Review Applied}\ }\textbf {\bibinfo {volume} {7}},\ \bibinfo
  {pages} {024019} (\bibinfo {year} {2017})}\BibitemShut {NoStop}%
\bibitem [{\citenamefont {Chui}\ \emph {et~al.}(2003)\citenamefont {Chui},
  \citenamefont {Hishinuma}, \citenamefont {Budakian}, \citenamefont {Mamin},
  \citenamefont {Kenny},\ and\ \citenamefont {Rugar}}]{chui2003}%
  \BibitemOpen
  \bibfield  {author} {\bibinfo {author} {\bibfnamefont {B.~W.}\ \bibnamefont
  {Chui}}, \bibinfo {author} {\bibfnamefont {Y.}~\bibnamefont {Hishinuma}},
  \bibinfo {author} {\bibfnamefont {R.}~\bibnamefont {Budakian}}, \bibinfo
  {author} {\bibfnamefont {H.~J.}\ \bibnamefont {Mamin}}, \bibinfo {author}
  {\bibfnamefont {T.~W.}\ \bibnamefont {Kenny}}, \ and\ \bibinfo {author}
  {\bibfnamefont {D.}~\bibnamefont {Rugar}},\ }in\ \href {\doibase
  10.1109/SENSOR.2003.1216966} {\emph {\bibinfo {booktitle} {TRANSDUCERS, The
  12th International Conference on Solid-State Sensors, Actuators and
  Microsystems}}},\ Vol.~\bibinfo {volume} {2}\ (\bibinfo {organization}
  {IEEE},\ \bibinfo {year} {2003})\ pp.\ \bibinfo {pages}
  {1120--1123}\BibitemShut {NoStop}%
\bibitem [{\citenamefont {Usenko}\ \emph {et~al.}(2011)\citenamefont {Usenko},
  \citenamefont {Vinante}, \citenamefont {Wijts},\ and\ \citenamefont
  {Oosterkamp}}]{usenko2011}%
  \BibitemOpen
  \bibfield  {author} {\bibinfo {author} {\bibfnamefont {O.}~\bibnamefont
  {Usenko}}, \bibinfo {author} {\bibfnamefont {A.}~\bibnamefont {Vinante}},
  \bibinfo {author} {\bibfnamefont {G.~H. C.~J.}\ \bibnamefont {Wijts}}, \ and\
  \bibinfo {author} {\bibfnamefont {T.~H.}\ \bibnamefont {Oosterkamp}},\ }\href
  {\doibase 10.1063/1.3570628} {\bibfield  {journal} {\bibinfo  {journal}
  {Applied Physics Letters}\ }\textbf {\bibinfo {volume} {98}},\ \bibinfo
  {pages} {133105} (\bibinfo {year} {2011})}\BibitemShut {NoStop}%
\bibitem [{Note1()}]{Note1}%
  \BibitemOpen
  \bibinfo {note} {Magnicon GMBH. Integrated 2-stage current sensor, type
  C70M116W}\BibitemShut {NoStop}%
\bibitem [{\citenamefont {Thoen}\ \emph {et~al.}(2017)\citenamefont {Thoen},
  \citenamefont {Bos}, \citenamefont {Haalebos}, \citenamefont {Klapwijk},
  \citenamefont {Baselmans},\ and\ \citenamefont {Endo}}]{thoen2017}%
  \BibitemOpen
  \bibfield  {author} {\bibinfo {author} {\bibfnamefont {D.~J.}\ \bibnamefont
  {Thoen}}, \bibinfo {author} {\bibfnamefont {B.~G.~C.}\ \bibnamefont {Bos}},
  \bibinfo {author} {\bibfnamefont {E.~A.~F.}\ \bibnamefont {Haalebos}},
  \bibinfo {author} {\bibfnamefont {T.~M.}\ \bibnamefont {Klapwijk}}, \bibinfo
  {author} {\bibfnamefont {J.~J.~A.}\ \bibnamefont {Baselmans}}, \ and\
  \bibinfo {author} {\bibfnamefont {A.}~\bibnamefont {Endo}},\ }\href {\doibase
  10.1109/TASC.2016.2631948} {\bibfield  {journal} {\bibinfo  {journal} {IEEE
  Transactions on Applied Superconductivity}\ } (\bibinfo {year} {2017}),\
  10.1109/TASC.2016.2631948}\BibitemShut {NoStop}%
\bibitem [{\citenamefont {Den~Haan}\ \emph {et~al.}(2014)\citenamefont
  {Den~Haan}, \citenamefont {Wijts}, \citenamefont {Galli}, \citenamefont
  {Usenko}, \citenamefont {Van~Baarle}, \citenamefont {Van Der~Zalm},\ and\
  \citenamefont {Oosterkamp}}]{haan2014}%
  \BibitemOpen
  \bibfield  {author} {\bibinfo {author} {\bibfnamefont {A.~M.~J.}\
  \bibnamefont {Den~Haan}}, \bibinfo {author} {\bibfnamefont {G.~H. C.~J.}\
  \bibnamefont {Wijts}}, \bibinfo {author} {\bibfnamefont {F.}~\bibnamefont
  {Galli}}, \bibinfo {author} {\bibfnamefont {O.}~\bibnamefont {Usenko}},
  \bibinfo {author} {\bibfnamefont {G.~J.~C.}\ \bibnamefont {Van~Baarle}},
  \bibinfo {author} {\bibfnamefont {D.~J.}\ \bibnamefont {Van Der~Zalm}}, \
  and\ \bibinfo {author} {\bibfnamefont {T.~H.}\ \bibnamefont {Oosterkamp}},\
  }\href {\doibase 10.1063/1.4868684} {\bibfield  {journal} {\bibinfo
  {journal} {Review of Scientific Instruments}\ }\textbf {\bibinfo {volume}
  {85}},\ \bibinfo {pages} {035112} (\bibinfo {year} {2014})}\BibitemShut
  {NoStop}%
\bibitem [{\citenamefont {Meyer}\ \emph {et~al.}(1989)\citenamefont {Meyer},
  \citenamefont {Silvera},\ and\ \citenamefont {Brandt}}]{meyer1989}%
  \BibitemOpen
  \bibfield  {author} {\bibinfo {author} {\bibfnamefont {E.~S.}\ \bibnamefont
  {Meyer}}, \bibinfo {author} {\bibfnamefont {I.~F.}\ \bibnamefont {Silvera}},
  \ and\ \bibinfo {author} {\bibfnamefont {B.~L.}\ \bibnamefont {Brandt}},\
  }\href@noop {} {\bibfield  {journal} {\bibinfo  {journal} {Review of
  Scientific Instruments}\ }\textbf {\bibinfo {volume} {60}},\ \bibinfo {pages}
  {2964} (\bibinfo {year} {1989})}\BibitemShut {NoStop}%
\bibitem [{\citenamefont {Korringa}(1950)}]{korringa1950}%
  \BibitemOpen
  \bibfield  {author} {\bibinfo {author} {\bibfnamefont {J.}~\bibnamefont
  {Korringa}},\ }\href {\doibase 10.1016/0031-8914(50)90105-4} {\bibfield
  {journal} {\bibinfo  {journal} {Physica}\ }\textbf {\bibinfo {volume} {16}},\
  \bibinfo {pages} {601} (\bibinfo {year} {1950})}\BibitemShut {NoStop}%
\bibitem [{\citenamefont {Lounasmaa}(1997)}]{lounasmaa1997}%
  \BibitemOpen
  \bibfield  {author} {\bibinfo {author} {\bibfnamefont {O.~V.}\ \bibnamefont
  {Lounasmaa}},\ }\href@noop {} {\bibfield  {journal} {\bibinfo  {journal}
  {Matematisk-Fysiske Meddelelser}\ ,\ \bibinfo {pages} {401}} (\bibinfo {year}
  {1997})}\BibitemShut {NoStop}%
\bibitem [{\citenamefont {Oja}\ and\ \citenamefont
  {Lounasmaa}(1997)}]{oja1997}%
  \BibitemOpen
  \bibfield  {author} {\bibinfo {author} {\bibfnamefont {A.~S.}\ \bibnamefont
  {Oja}}\ and\ \bibinfo {author} {\bibfnamefont {O.~V.}\ \bibnamefont
  {Lounasmaa}},\ }\href {\doibase 10.1103/RevModPhys.69.1} {\bibfield
  {journal} {\bibinfo  {journal} {Reviews of Modern Physics}\ }\textbf
  {\bibinfo {volume} {69}},\ \bibinfo {pages} {1} (\bibinfo {year}
  {1997})}\BibitemShut {NoStop}%
\bibitem [{\citenamefont {Pobell}(2007)}]{pobell2007}%
  \BibitemOpen
  \bibfield  {author} {\bibinfo {author} {\bibfnamefont {F.}~\bibnamefont
  {Pobell}},\ }\href@noop {} {\emph {\bibinfo {title} {Matter and Methods at
  Low Temperatures}}}\ (\bibinfo  {publisher} {Springer Science \& Business
  Media},\ \bibinfo {year} {2007})\BibitemShut {NoStop}%
\bibitem [{\citenamefont {Isaac}\ \emph {et~al.}(2016)\citenamefont {Isaac},
  \citenamefont {Gleave}, \citenamefont {Nasr}, \citenamefont {Nguyen},
  \citenamefont {Curley}, \citenamefont {Yoder}, \citenamefont {Moore},
  \citenamefont {Chen},\ and\ \citenamefont {Marohn}}]{isaac2016}%
  \BibitemOpen
  \bibfield  {author} {\bibinfo {author} {\bibfnamefont {C.~E.}\ \bibnamefont
  {Isaac}}, \bibinfo {author} {\bibfnamefont {C.~M.}\ \bibnamefont {Gleave}},
  \bibinfo {author} {\bibfnamefont {P.~T.}\ \bibnamefont {Nasr}}, \bibinfo
  {author} {\bibfnamefont {H.~L.}\ \bibnamefont {Nguyen}}, \bibinfo {author}
  {\bibfnamefont {E.~A.}\ \bibnamefont {Curley}}, \bibinfo {author}
  {\bibfnamefont {J.~L.}\ \bibnamefont {Yoder}}, \bibinfo {author}
  {\bibfnamefont {E.~W.}\ \bibnamefont {Moore}}, \bibinfo {author}
  {\bibfnamefont {L.}~\bibnamefont {Chen}}, \ and\ \bibinfo {author}
  {\bibfnamefont {J.~A.}\ \bibnamefont {Marohn}},\ }\href {\doibase
  10.1063/1.2833582} {\bibfield  {journal} {\bibinfo  {journal} {Physical
  Chemistry Chemical Physics}\ }\textbf {\bibinfo {volume} {18}},\ \bibinfo
  {pages} {8806} (\bibinfo {year} {2016})}\BibitemShut {NoStop}%
\bibitem [{\citenamefont {Abragam}(1961)}]{abragam1961}%
  \BibitemOpen
  \bibfield  {author} {\bibinfo {author} {\bibfnamefont {A.}~\bibnamefont
  {Abragam}},\ }\href@noop {} {\emph {\bibinfo {title} {Principles of nuclear
  magnetism (International series of monographs on physics)}}}\ (\bibinfo
  {publisher} {Clarendon Press, Oxford},\ \bibinfo {year} {1961})\ p.\ \bibinfo
  {pages} {128}\BibitemShut {NoStop}%
\bibitem [{\citenamefont {Bloch}(1946)}]{bloch1946}%
  \BibitemOpen
  \bibfield  {author} {\bibinfo {author} {\bibfnamefont {F.}~\bibnamefont
  {Bloch}},\ }\href {\doibase 10.1103/PhysRev.70.460} {\bibfield  {journal}
  {\bibinfo  {journal} {Physical review}\ }\textbf {\bibinfo {volume} {70}},\
  \bibinfo {pages} {460} (\bibinfo {year} {1946})}\BibitemShut {NoStop}%
\bibitem [{\citenamefont {Slichter}(1990)}]{slichter1990}%
  \BibitemOpen
  \bibfield  {author} {\bibinfo {author} {\bibfnamefont {C.~P.}\ \bibnamefont
  {Slichter}},\ }\href@noop {} {\emph {\bibinfo {title} {Principles of Magnetic
  Resonance, volume 1 of Springer Series in Solid-State Sciences}}}\ (\bibinfo
  {publisher} {Springer Berlin Heidelberg, Berlin, Heidelberg},\ \bibinfo
  {year} {1990})\BibitemShut {NoStop}%
\bibitem [{\citenamefont {Mulkern}\ and\ \citenamefont
  {Williams}(1993)}]{mulkern1993}%
  \BibitemOpen
  \bibfield  {author} {\bibinfo {author} {\bibfnamefont {R.~V.}\ \bibnamefont
  {Mulkern}}\ and\ \bibinfo {author} {\bibfnamefont {M.~L.}\ \bibnamefont
  {Williams}},\ }\href {\doibase 10.1118/1.597063} {\bibfield  {journal}
  {\bibinfo  {journal} {Medical physics}\ }\textbf {\bibinfo {volume} {20}},\
  \bibinfo {pages} {5} (\bibinfo {year} {1993})}\BibitemShut {NoStop}%
\bibitem [{\citenamefont {Murase}\ and\ \citenamefont
  {Tanki}(2011)}]{murase2011}%
  \BibitemOpen
  \bibfield  {author} {\bibinfo {author} {\bibfnamefont {K.}~\bibnamefont
  {Murase}}\ and\ \bibinfo {author} {\bibfnamefont {N.}~\bibnamefont {Tanki}},\
  }\href {\doibase 10.1016/j.mri.2010.07.003} {\bibfield  {journal} {\bibinfo
  {journal} {Magnetic resonance imaging}\ }\textbf {\bibinfo {volume} {29}},\
  \bibinfo {pages} {126} (\bibinfo {year} {2011})}\BibitemShut {NoStop}%
\bibitem [{\citenamefont {De~Voogd}\ \emph {et~al.}(2017)\citenamefont
  {De~Voogd}, \citenamefont {Wagenaar},\ and\ \citenamefont
  {Oosterkamp}}]{voogd2017}%
  \BibitemOpen
  \bibfield  {author} {\bibinfo {author} {\bibfnamefont {J.~M.}\ \bibnamefont
  {De~Voogd}}, \bibinfo {author} {\bibfnamefont {J.~J.~T.}\ \bibnamefont
  {Wagenaar}}, \ and\ \bibinfo {author} {\bibfnamefont {T.~H.}\ \bibnamefont
  {Oosterkamp}},\ }\href {\doibase 10.1038/srep42239} {\bibfield  {journal}
  {\bibinfo  {journal} {Scientific Reports}\ }\textbf {\bibinfo {volume} {7}},\
  \bibinfo {pages} {42239} (\bibinfo {year} {2017})}\BibitemShut {NoStop}%
\bibitem [{\citenamefont {Herzog}\ \emph {et~al.}(2014)\citenamefont {Herzog},
  \citenamefont {Cadeddu}, \citenamefont {Xue}, \citenamefont {Peddibhotla},\
  and\ \citenamefont {Poggio}}]{herzog2014}%
  \BibitemOpen
  \bibfield  {author} {\bibinfo {author} {\bibfnamefont {B.~E.}\ \bibnamefont
  {Herzog}}, \bibinfo {author} {\bibfnamefont {D.}~\bibnamefont {Cadeddu}},
  \bibinfo {author} {\bibfnamefont {F.}~\bibnamefont {Xue}}, \bibinfo {author}
  {\bibfnamefont {P.}~\bibnamefont {Peddibhotla}}, \ and\ \bibinfo {author}
  {\bibfnamefont {M.}~\bibnamefont {Poggio}},\ }\href {\doibase
  10.1063/1.4892361} {\bibfield  {journal} {\bibinfo  {journal} {Applied
  physics letters}\ }\textbf {\bibinfo {volume} {105}},\ \bibinfo {pages}
  {043112} (\bibinfo {year} {2014})}\BibitemShut {NoStop}%
\bibitem [{\citenamefont {Yazdanian}\ \emph {et~al.}(2008)\citenamefont
  {Yazdanian}, \citenamefont {Marohn},\ and\ \citenamefont
  {Loring}}]{yazdanian2008}%
  \BibitemOpen
  \bibfield  {author} {\bibinfo {author} {\bibfnamefont {S.~M.}\ \bibnamefont
  {Yazdanian}}, \bibinfo {author} {\bibfnamefont {J.~A.}\ \bibnamefont
  {Marohn}}, \ and\ \bibinfo {author} {\bibfnamefont {R.~F.}\ \bibnamefont
  {Loring}},\ }\href {\doibase 10.1063/1.2932254} {\bibfield  {journal}
  {\bibinfo  {journal} {The Journal of chemical physics}\ }\textbf {\bibinfo
  {volume} {128}},\ \bibinfo {pages} {224706} (\bibinfo {year}
  {2008})}\BibitemShut {NoStop}%
\bibitem [{\citenamefont {Hoepker}\ \emph {et~al.}(2011)\citenamefont
  {Hoepker}, \citenamefont {Lekkala}, \citenamefont {Loring},\ and\
  \citenamefont {Marohn}}]{hoepker2011}%
  \BibitemOpen
  \bibfield  {author} {\bibinfo {author} {\bibfnamefont {N.}~\bibnamefont
  {Hoepker}}, \bibinfo {author} {\bibfnamefont {S.}~\bibnamefont {Lekkala}},
  \bibinfo {author} {\bibfnamefont {R.~F.}\ \bibnamefont {Loring}}, \ and\
  \bibinfo {author} {\bibfnamefont {J.~A.}\ \bibnamefont {Marohn}},\ }\href
  {\doibase 10.1021/jp207387d} {\bibfield  {journal} {\bibinfo  {journal} {The
  Journal of Physical Chemistry B}\ }\textbf {\bibinfo {volume} {115}},\
  \bibinfo {pages} {14493} (\bibinfo {year} {2011})}\BibitemShut {NoStop}%
\bibitem [{\citenamefont {Yazdanian}\ \emph {et~al.}(2009)\citenamefont
  {Yazdanian}, \citenamefont {Hoepker}, \citenamefont {Kuehn}, \citenamefont
  {Loring},\ and\ \citenamefont {Marohn}}]{yazdanian2009}%
  \BibitemOpen
  \bibfield  {author} {\bibinfo {author} {\bibfnamefont {S.~M.}\ \bibnamefont
  {Yazdanian}}, \bibinfo {author} {\bibfnamefont {N.}~\bibnamefont {Hoepker}},
  \bibinfo {author} {\bibfnamefont {S.}~\bibnamefont {Kuehn}}, \bibinfo
  {author} {\bibfnamefont {R.~F.}\ \bibnamefont {Loring}}, \ and\ \bibinfo
  {author} {\bibfnamefont {J.~A.}\ \bibnamefont {Marohn}},\ }\href {\doibase
  10.1021/nl9004332} {\bibfield  {journal} {\bibinfo  {journal} {Nano letters}\
  }\textbf {\bibinfo {volume} {9}},\ \bibinfo {pages} {2273} (\bibinfo {year}
  {2009})}\BibitemShut {NoStop}%
\bibitem [{\citenamefont {Moore}\ \emph {et~al.}(2009)\citenamefont {Moore},
  \citenamefont {Lee}, \citenamefont {Hickman}, \citenamefont {Wright},
  \citenamefont {Harrell}, \citenamefont {Borbat}, \citenamefont {Freed},\ and\
  \citenamefont {Marohn}}]{moore2009}%
  \BibitemOpen
  \bibfield  {author} {\bibinfo {author} {\bibfnamefont {E.~W.}\ \bibnamefont
  {Moore}}, \bibinfo {author} {\bibfnamefont {S.}~\bibnamefont {Lee}}, \bibinfo
  {author} {\bibfnamefont {S.~A.}\ \bibnamefont {Hickman}}, \bibinfo {author}
  {\bibfnamefont {S.~J.}\ \bibnamefont {Wright}}, \bibinfo {author}
  {\bibfnamefont {L.~E.}\ \bibnamefont {Harrell}}, \bibinfo {author}
  {\bibfnamefont {P.~P.}\ \bibnamefont {Borbat}}, \bibinfo {author}
  {\bibfnamefont {J.~H.}\ \bibnamefont {Freed}}, \ and\ \bibinfo {author}
  {\bibfnamefont {J.~A.}\ \bibnamefont {Marohn}},\ }\href {\doibase
  10.1073/pnas.0908120106} {\bibfield  {journal} {\bibinfo  {journal}
  {Proceedings of the National Academy of Sciences}\ }\textbf {\bibinfo
  {volume} {106}},\ \bibinfo {pages} {22251} (\bibinfo {year}
  {2009})}\BibitemShut {NoStop}%
\bibitem [{\citenamefont {Oosterkamp}\ \emph {et~al.}(2010)\citenamefont
  {Oosterkamp}, \citenamefont {Poggio}, \citenamefont {Degen}, \citenamefont
  {Mamin},\ and\ \citenamefont {Rugar}}]{oosterkamp2010}%
  \BibitemOpen
  \bibfield  {author} {\bibinfo {author} {\bibfnamefont {T.~H.}\ \bibnamefont
  {Oosterkamp}}, \bibinfo {author} {\bibfnamefont {M.}~\bibnamefont {Poggio}},
  \bibinfo {author} {\bibfnamefont {C.~L.}\ \bibnamefont {Degen}}, \bibinfo
  {author} {\bibfnamefont {H.~J.}\ \bibnamefont {Mamin}}, \ and\ \bibinfo
  {author} {\bibfnamefont {D.}~\bibnamefont {Rugar}},\ }\href@noop {}
  {\bibfield  {journal} {\bibinfo  {journal} {Applied Physics Letters}\
  }\textbf {\bibinfo {volume} {96}},\ \bibinfo {pages} {083107} (\bibinfo
  {year} {2010})}\BibitemShut {NoStop}%
\bibitem [{\citenamefont {Moores}\ \emph {et~al.}(2015)\citenamefont {Moores},
  \citenamefont {Eichler}, \citenamefont {Tao}, \citenamefont {Takahashi},
  \citenamefont {Navaretti},\ and\ \citenamefont {Degen}}]{moores2015}%
  \BibitemOpen
  \bibfield  {author} {\bibinfo {author} {\bibfnamefont {B.~A.}\ \bibnamefont
  {Moores}}, \bibinfo {author} {\bibfnamefont {A.}~\bibnamefont {Eichler}},
  \bibinfo {author} {\bibfnamefont {Y.}~\bibnamefont {Tao}}, \bibinfo {author}
  {\bibfnamefont {H.}~\bibnamefont {Takahashi}}, \bibinfo {author}
  {\bibfnamefont {P.}~\bibnamefont {Navaretti}}, \ and\ \bibinfo {author}
  {\bibfnamefont {C.~L.}\ \bibnamefont {Degen}},\ }\href@noop {} {\bibfield
  {journal} {\bibinfo  {journal} {Applied Physics Letters}\ }\textbf {\bibinfo
  {volume} {106}},\ \bibinfo {pages} {213101} (\bibinfo {year}
  {2015})}\BibitemShut {NoStop}%
\bibitem [{\citenamefont {Chen}\ \emph {et~al.}(2013)\citenamefont {Chen},
  \citenamefont {Longenecker}, \citenamefont {Moore},\ and\ \citenamefont
  {Marohn}}]{chen2013}%
  \BibitemOpen
  \bibfield  {author} {\bibinfo {author} {\bibfnamefont {L.}~\bibnamefont
  {Chen}}, \bibinfo {author} {\bibfnamefont {J.~G.}\ \bibnamefont
  {Longenecker}}, \bibinfo {author} {\bibfnamefont {E.~W.}\ \bibnamefont
  {Moore}}, \ and\ \bibinfo {author} {\bibfnamefont {J.~A.}\ \bibnamefont
  {Marohn}},\ }\href {\doibase 10.1063/1.4795018} {\bibfield  {journal}
  {\bibinfo  {journal} {Applied physics letters}\ }\textbf {\bibinfo {volume}
  {102}},\ \bibinfo {pages} {132404} (\bibinfo {year} {2013})}\BibitemShut
  {NoStop}%
\bibitem [{\citenamefont {Grapengeter}\ \emph {et~al.}(1980)\citenamefont
  {Grapengeter}, \citenamefont {Kosfeld},\ and\ \citenamefont
  {Offergeld}}]{grapengeter1980}%
  \BibitemOpen
  \bibfield  {author} {\bibinfo {author} {\bibfnamefont {H.~H.}\ \bibnamefont
  {Grapengeter}}, \bibinfo {author} {\bibfnamefont {R.}~\bibnamefont
  {Kosfeld}}, \ and\ \bibinfo {author} {\bibfnamefont {H.~W.}\ \bibnamefont
  {Offergeld}},\ }\href {\doibase 10.1016/0032-3861(80)90304-3} {\bibfield
  {journal} {\bibinfo  {journal} {Polymer}\ }\textbf {\bibinfo {volume} {21}},\
  \bibinfo {pages} {829} (\bibinfo {year} {1980})}\BibitemShut {NoStop}%
\end{thebibliography}%
\bibliographystyle{apsrev4-1}

\end{document}